\documentclass[a4paper,fleqn,usenatbib]{mnras}

\usepackage{newtxtext,newtxmath}

\usepackage[T1]{fontenc}
\usepackage{ae,aecompl}

\usepackage[utf8]{inputenc} 
\usepackage{lipsum}
\usepackage{graphicx}	
\usepackage{amsmath}	
\usepackage{amssymb}	

\title[Morphology of H$\alpha$ emission in CALIFA galaxies]{The Morphology of H$\alpha$ emission in CALIFA galaxies}

\author[Patricia Novais et al.]{
P.~M.~Novais$^{1}$\thanks{E-mail: patynovais@gmail.com} \&
L. Sodré Jr.$^{1}$
\\
$^{1}$Universidade de São Paulo; Instituto de Astronomia, Geofísica e Ciências Atmosféricas, Rua do Matão 1226, 05508-090 Sâo Paulo, Brazil
}

\date{Accepted XXX. Received YYY; in original form ZZZ}

\pubyear{2018}

\begin{document}
\label{firstpage}
\pagerange{\pageref{firstpage}--\pageref{lastpage}}
\maketitle

\begin{abstract}
We have determined the H$\alpha$ emission line radial profiles for a sample of 86 face-on galaxies observed in the CALIFA survey and analyzed with the Pipe3D pipeline. From a visual analysis  we propose a two step classification of these profiles. Initially, they were divided in two classes with respect to the maximum of the  H$\alpha$ emission: $C$ (for central) profiles have the maximum line emission at the galaxy centre, whereas $EX$ (for extended) profiles have the maximum  H$\alpha$ emission outside the galaxy center. After, we divided the $C$ galaxies in two classes, $CE$ and $CL$ (where E and L stands here for 'early' and 'late'), through the value of $c_r$, the concentration index in the $r$-band. We analyzed the profile class dependence of several galaxy parameters, as well as examined the nature of line emission through the BPT diagram. We notice that almost $75$\% of the sample is in the $C$ class. Elliptical and S0 galaxies dominate the $CE$ class, with  spiral galaxies being found mostly in the $CL$  and $EX$  classes. We also notice that spirals in each of these classes have different properties, with $CL$ objects seeming less evolved than those in the $EX$ class.  
\end{abstract}




\section{Introduction}
\label{sec:intro}
The study of emission lines provides a powerful way to investigate a large variety of physical processes ocurring in galaxies. This is particularly true for H$\alpha$, which is often one of the most prominent lines in the optical interval. Indeed, star-formation may produce a powerful  H$\alpha$ line emission from recombination in nebular regions ionized by the radiation field of young, massive stars \citep[e.g,][]{osterbrock2005}, and the flux of this line is often used as an indicator of the star formation rate \citep{kennicutt1998_SFR,starlightI,Calzetti2013,madau2014}. But other sources than star-formation can also ionize the gas and lead to H$\alpha$ emission, like nuclear activity driven by AGNs, where the ionizing photons are produced in an accretion disk around a massive black hole \citep[][]{veron-cetty-veron2000}, or shocks in the gas \citep[][]{dopita1995,alatalo2016}.

In some cases the origin of the emission is not clear, as is the case of the LINERS. This terminology was introduced by \cite{heckman1980} for objects with emission lines with widths similar to that found in Seyfert galaxies but showing lower excitation when compared to those galaxies. Later, \cite{trinchieri1991} and \cite{binette1994} proposed that, instead, the LINER emission could originate through ionization by evolved stars, like post-AGB stars or white dwarfs. \cite{cidfernandes2010} have argued that many galaxies with weak LINER-like emission lines could be {\it retired} galaxies, where the star formation ceased and which are now being ionized by old stars. The introduction of integral field studies helped to clarify this issue by showing that the LINER emission in some galaxies is actually extended and can not be produced by nuclear activity alone \citep[e.g. ][]{singh2013,ricci2014,belfiore2016}.

Many studies on the morphology of the H$\alpha$ emission were carried out aiming to derive structural or kinematical properties of the galaxies or of the gas emission. \cite{karachentsev2010} investigated the H$\alpha$ emission for the Local Volume, finding evidence of stellar formation in early-type objects both in knots and in the central regions, interpreting it as indication of accretion of intergalactic gas. \cite{nelson2012} compared spatially resolved maps in H$\alpha$ with maps of the continuum in the R band for a sample of $z\sim1$ star forming galaxies, showing that although both the maps are quite similar,  H$\alpha$ emission is more extended and more clumpy than the continuous emission. \cite{gavazzi2018}, analyzed a sample of 147 early-type galaxies (ETG) from the ATLAS$^{3D}$ survey \citep{Cappellari2011} and found that 37\% of them exhibited some emission in H$\alpha$, with about 14\% of them exhibiting relatively strong emission, produced mainly by low mass S0 with stellar and gaseous fast rotating discs.

Since H$\alpha$ emission (and its luminosity) is an important proxy for both nuclear activity and formation of stars, spatially resolved maps of H$\alpha$  may provide important informations about ionization mechanisms, star formation rates, mass assembly, and other galaxy properties.  The advent of integral field spectroscopy surveys, like ATLAS$^{3D}$ \citep{Cappellari2011},  SAMI Galaxy Survey \citep{croom2012}, CALIFA \citep{sanchez2012}, and MaNGA \citep{bundy2015}, enabled the investigation of the spatially resolved properties of galaxies and their relation to processes influencing the formation and evolution of galaxies, like quenching. With spatially resolved data it is possible to study the  radial dependence of star formation histories \citep{perez2013,garciabenito2017,rosa2017,goddard2017, amorim2017} or evaluate star formation rates inside galaxies, both locally and globally \citep{canodiaz2016,rosa2016,kokusho2017,medling2018}. 

The morphology of the H$\alpha$ has often being addressed via the distribution within a galaxy of the star formation rate or other star formation indicators. As part of a series of studies on the star formation morphology in galaxies of the Virgo Cluster, \cite{Koopmann2001} analyzed R-band and H$\alpha$ images, radial profiles, integrated fluxes, and concentration indices to establish links between the H$\alpha$ morphologies of Virgo Cluster spiral galaxies and the types of interaction(s) that may have affected them. These authors proposed a set of   classes (combinations of normal, enhanced, anemic, and truncated) to describe the global galaxy star formation morphology. Analyzing a small sample  of galaxies in the MaNGA prototype run  (P-MaNGA), \cite{li2015} classified the radial profiles of their three diagnostic parameters (Dn4000), EW(H$\delta$), and EW(H$\alpha$)) as either ‘centrally quiescent’  or ‘centrally star forming’. Recently, \cite{spindler2018} investigated the spatial distribution of star formation in a large sample of galaxies of the MaNGA survey \citep{bundy2015}, revealing the existence of two groups of galaxies, which they named 'Centrally Suppressed' and 'Unsuppressed' regarding the radial distribution of their specific star formation rates. Analogously, \cite{medling2018} studied the resolved star formation of a sample of galaxies in the SAMI survey, also finding a split between galaxy populations with centrally concentrated star formation and those with flatter star formation profiles.

In this paper we consider the classification of H$\alpha$ emission radial profiles for a sample of 86 almost face-on  CALIFA survey galaxies. We focus on H$\alpha$ to avoid, in the classification stage, discriminating between emission due to star formation, AGN or old stars. We propose an objective procedure where we first divide galaxies in two classes -  those where the peak of the H$\alpha$ emission is in the galaxy center and those where this peak is outside the center- and,  after, we divide the galaxies in the first group in two by using the light concentration in the $r$-band to take in to account in this classification the bimodal nature of the galaxy population. Indeed, in the local universe, a bimodality is present in many galaxy properties, such as colors and stellar formation rates \citep{Strateva2001, omill2006, baldry2006, zibetti2017, Nelson2018}, by which they  can be separated into two large groups, early-types and late-types. Early-types comprise essentially elliptical and lenticular galaxies with old stellar populations, red colors, high masses and almost none star formation \citep{Trager2000, conselice2006, vanDokkum2007}. In contrast, late-type galaxies are mostly spirals (with and without bars) and irregulars galaxies,  characterized by younger populations, blue colors, lower stellar masses and active star formation, when compared to early-type galaxies \citep{Bell2003,abilio2006}.  We also investigate whether our profile classification is correlated with other galaxy parameters or properties, in particular those inferred from population synthesis. The analysis of  stellar populations through galaxy spectral synthesis \citep[e.g., ][]{tinsley76,tinsley80,leitherer99,bruzualecharlot2003,starlightI}  allows to determine properties such as mean stellar ages, metallicity, stellar mass and star formation history that are useful to characterize the galaxies. Additionally, morphological types, brightness profiles \citep{deVaucouleurs1948,Sersic1963} and also concentration parameters have been shown to relate to many physical features of galaxies or of their galaxy populations \citep{okamura1984,graham2001,gini2003,conselice2014}. \cite{Strateva2001}.

This paper begins with a description of our sample in Section \ref{sec:Data}. There we describe some relevant details of the CALIFA Survey, the Pipe3D pipeline and how we deal with the dust extinction. In Section \ref{sec:analysis} we explain the structural and physical parameters used here, as well as the procedure to determine and classify the radial H$\alpha$ profiles. Section \ref{sec:results} present a comparative study between the profile classes and other physical parameters. In Section \ref{sec:discussion} we discuss these results, making clear the relations between profile classes and galaxy properties. Section \ref{sec:summary} summarizes the work presented here.


\section{Data}
\label{sec:Data}

The galaxies analyzed in this work were selected from the CALIFA Survey, with datacubes treated by the Pipe3D pipeline, both briefly described below. Additionally, we present our procedure to correct the H$\alpha$ emission for dust attenuation, as well as the criteria adopted to select the final sample.

\subsection{The CALIFA sample}
\label{sec:califa}

The Calar Alto Legacy  Integral Field Area \citep*[CALIFA;][]{califa_surveyI} survey, in its  second release \citep{califadr2}, observed 200 galaxies of all Hubble types in the nearby universe. The galaxy sample of this survey was selected from the SDSS-DR7 photometric catalogue \citep{sdssDR7}, with constraints on isophotal diameter ($45" < D_{25} < 80"$) and on the redshift range ($0.005 < z < 0.03$).

The objects were observed with the integral-field spectrograph
PMAS/PPak mounted on the 3.5 m telescope at the Calar Alto observatory and with two different spectral setups for each galaxy: a V500 grating with low-resolution ($\lambda/\Delta\lambda=850$ in 5000Å) in the wavelength range 3745Å-7500Å, and a V1200 grating with medium-resolution ($\lambda/\Delta\lambda=1650$ in 4500Å) in the wavelength range 3400Å-4750Å. A more complete description of the CALIFA survey can be found in \cite{walcher2014}.

\begin{figure*}
\includegraphics[width=0.49 \columnwidth,angle=0]{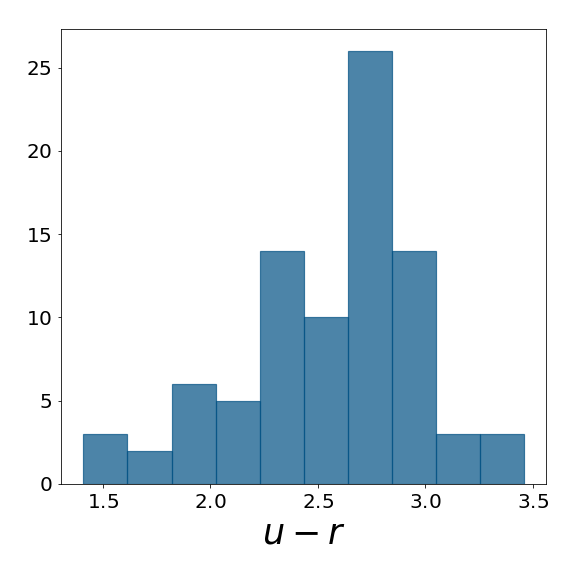}
\includegraphics[width=0.49 \columnwidth,angle=0]{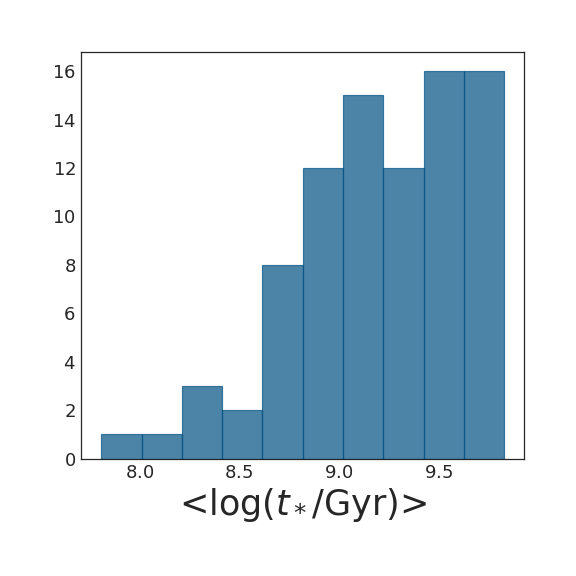}
\includegraphics[width=0.49 \columnwidth,angle=0]{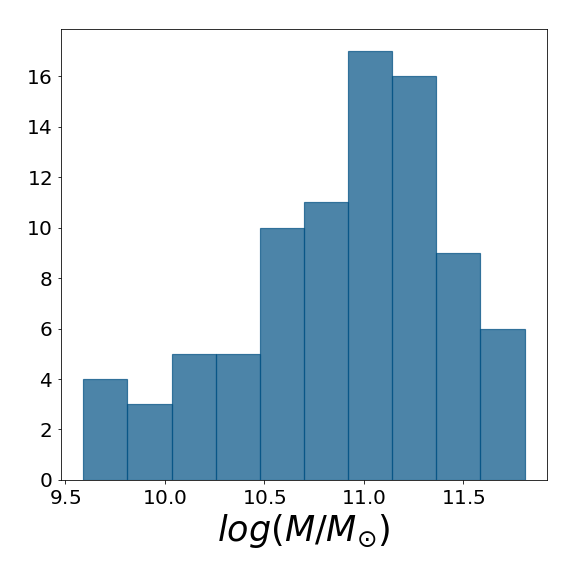}
\includegraphics[width=0.49 \columnwidth,angle=0]{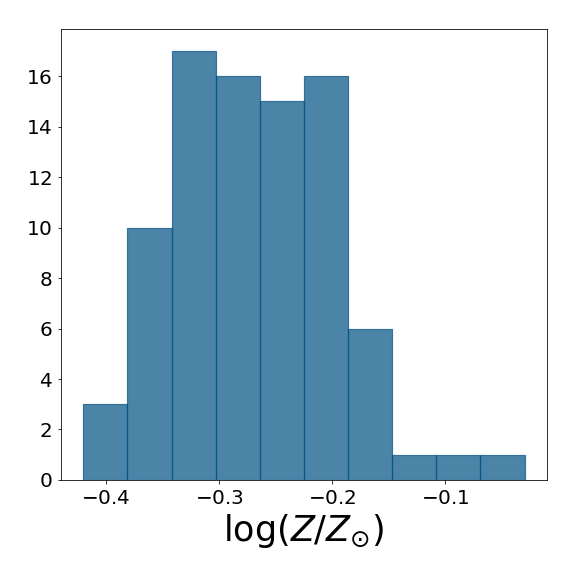}
\caption[]{Some properties of the galaxy sample discussed in this paper, comprised of face-on galaxies  from CALIFA/DR2.}
\label{distrib_propriedades}
\end{figure*}

\subsection{Pipe3D}
Pipe3D \citep{Pipe3d,I_Pipe3d} is an analysis pipeline developed to extract information from IFS datacubes of surveys like CALIFA, MaNGA and SAMI \citep{califa_surveyI, bundy2015, croom2012}. 

Pipe3D depends on a spectral synthesis tool \citep[FIT3D][]{I_Pipe3d} and comprises the GSD156 library of simple stellar populations \citep{cid2013}, with 156 templates covering 39 stellar ages (from 1Myr to 14.1Gyr), and 4 metallicities ($Z/Z_{\odot{}}=0.2, 0.4, 1, \text{and} 1.5$). All the steps used to obtain  the dataproducts are well explained in \cite{I_Pipe3d,Pipe3d}.

In this work we have started by using the results of Pipe3D for 200 galaxies from CALIFA/DR2 \citep{califadr2}, with the V500 setup. From the many dataproducts obtained with Pipe3D, we used in our analysis the H$\beta$, [OIII]$\lambda$5007, H$\alpha$ and [NII]$\lambda$6583 emission lines maps, all of them in units of $10^{-16}$erg s$^{-1}$ cm$^{-2}$.

\subsection{H$\alpha$ maps: source and dust attenuation}
\label{sec:maps} 
Most of our analysis is based on H$\alpha$ maps. To take in to account the effect of extiction due to dust within the galaxy on this and other line intensities, we corrected the dust attenuation by using a  screen model \citep[e.g.][]{calzetti1994}. We adopted the extinction law of \cite{cardelli1989}, with $R_V=3.1$. These assumptions lead to a relation between the extinction coefficient in the V-band and the H$\alpha$/H$\beta$ intensity ratio as:
\begin{equation}
A_V = 6.10 \times \log \Bigg[\frac{H\alpha}{H\beta}\Bigg] - 2.78. 
\end{equation}
In order to correct the dust attenuation pixel by pixel, we have used the mean value of $A_V$, calculated  in $r > 0.20 a$ (where $a$ is the semi-major axis of the galaxy image; see Sect. \ref{sec:analysis}), to avoid contamination of this estimate by non-stellar nuclear activity.

\subsection{Our sample}
From the 200 galaxies observed in CALIFA/DR2, we selected nearly face-on objects, i.e., those with ellipticity $\epsilon < 0.35$ in the V band (see next section). Our final sample contains 86 galaxies of different morphological types, which we divided in three morphological classes as shown in Table  \ref{tabela_morf}. This sample covers a range of physical properties typical of those found in bright galaxies in the local universe: it has colors in the interval $1.4 < u-r < 3.4$, mean stellar ages in $\sim 10^{7.8} < t_* ~({\rm yr}) <10^{9.8}$, stellar masses in $\sim 10^{9.5} < M_*/M_{\odot} < 10^{11.8}$ and metallicities in $-0.4 < \log(Z/Z_\odot) < -0.03$. Figure \ref{distrib_propriedades} shows the distribution of these properties for the 86 galaxies in the sample. Table \ref{tab01}, in the Appendix, summarizes relevant properties for each galaxy.


\section{Analysis}
\label{sec:analysis} 

In this section we describe some morphological estimators for the $V$-band emission and our definitions of the concentration of the $H\alpha$ emission. We also show  how we measured the $H\alpha$ emission profiles and present a procedure for their classification.

\subsection{Morphological and Structural Parameters}

\begin{table}
\vspace{0.3cm}
\centering
\begin{tabular}{|c| c| c|}
\hline 
Morph. Class & Hubble type & \# \\ 
\hline \hline                
E\underline{ }S0 & E0-E7, S0        & 30 \\
S\underline{ }early & Sa, Sab, Sb          & 23 \\ 
S\underline{ }late &  Sbc, Sc, Scd, Sd & 33 \\ [1ex] 
\hline 
\end{tabular}
\caption[]{Morphological classes of our sample, with the respective Hubble Types and number of objects.}
\label{tabela_morf} 
\end{table}

\begin{figure*}
\begin{center}
\includegraphics[width=0.49 \columnwidth,angle=0]{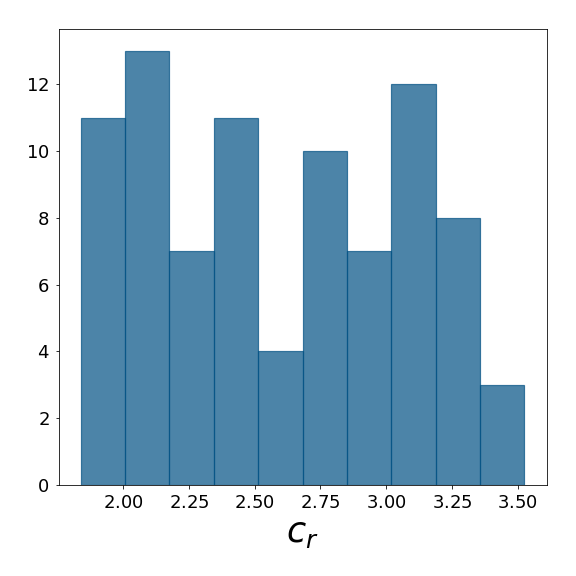}
\includegraphics[width=0.49 \columnwidth,angle=0]{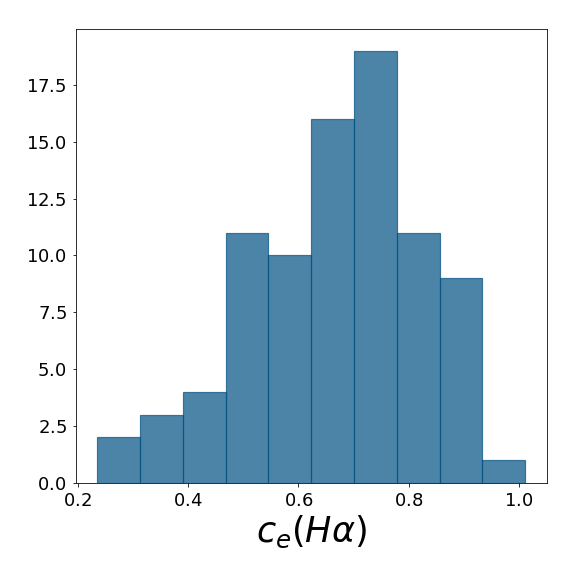}
\includegraphics[width=0.49 \columnwidth,angle=0]{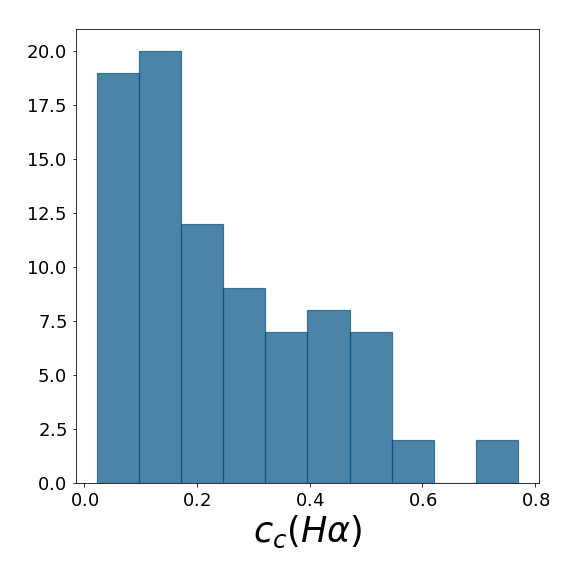}
\caption[]{Distribution of the concentrations $c_r$, $c_e(H\alpha)$ and $c_c(H\alpha)$ for the galaxies in the sample.}
\label{figure: distribuicoes_concentracao}
\end{center}
\end{figure*}

We have determined quantitative morphological parameters for the images delivered by Pipe3D for the galaxies in our sample through their image moments $\mu_{pq}$. The raw and  central image moments of order $(p+q)$ are given by \citep{flusser09} 

\begin{equation}
M_{pq}= \sum_{i=1}^{n_x} \sum_{j=1}^{n_y} x_i^p y_j^{q} F(x_i,y_j)
\end{equation}
and
\begin{equation}
\mu_{pq}= \sum_{i=1}^{n_x} \sum_{j=1}^{n_y} (x_i - \bar{x})^p (y_j - \bar{y})^{q} F(x_i,y_j),
\end{equation}
where $(x_i,y_i)$ are the coordinates of the $i$-th pixel of the image, ($\bar{x},\bar{y}$) is the centroid of the emission, given by
\[\bar{x} = \frac{M_{10}}{M_{00}} ~~~~~~~~~~
\bar{y} = \frac{M_{01}}{M_{00}},\] 
and $F(x_i,y_i)$ is the flux in that pixel.

Image moments are useful for obtaining the major and minor semi-axes \textit{a} and \textit{b}, the ellipticity $\epsilon$, the inclination angle $\theta$ and the mean radius $\bar{R}$ of the ellipse that fits the image of a galaxy. We apply the moments of the image in the V band map reconstructed by Pipe3D of each galaxy, so that the flux $F(x_i,y_i)$ and all the derived (ellipse) parameters are in $V$ band. The moments were calculated using  all pixels with positive fluxes.

Since we have galaxies of different sizes, it is convenient to normalize the scales (and in particular the major semi-axis) of each galaxy by its effective radius, $R_e(V)$, which is defined as the radius which comprises half of the $V$-band galaxy's flux.

To help in the discussion on the structural properties of the H$\alpha$ emission for each galaxy, we have derived two concentration parameters. The first is the effective concentration, $c_e(H\alpha)$, which measures the ratio of the radius containing 50\% of the H$\alpha$ emission ($R_e(H\alpha)$) in the H$\alpha$ maps, to the major semi-axis obtained in the $V$ band, $a$, through the image moments, and, second, the central concentration, $c_c(H\alpha)$, which measures the ratio between the H$\alpha$ flux within $0.2a$ and the total  H$\alpha$ flux: 
\begin{equation}
c_{e}(H\alpha) = \frac{R_{e}(H\alpha)}{a}
\end{equation}
\begin{equation}
c_c(H\alpha) = \frac{I_{H\alpha}(0.2a)}{I_{H\alpha}(a)}
\end{equation}
The concentraction $c_r$ of the luminosity in the r band, as defined by \cite{Strateva2001}, was also calculated for the galaxies in the sample, and the distributions of these three concentrations are shown in Figure \ref{figure: distribuicoes_concentracao}. The concentration $c_r$ was used by \cite{Strateva2001} to investigate the bimodality of the galaxy population (which is also present in the figure), and we show below that it is also useful to analyze H$\alpha$ profiles.

\subsection{Radial Profiles of H$\alpha$}
\label{sec:profiles}

In this section we discuss, initially, our procedure to obtain radial H$\alpha$ profiles and, after, the criteria we adopted to classify them in to three classes.

Given the H$\alpha$ image of a galaxy, we first define its centroid and other geometric parameters as the same obtained by using the moments of the image in the $V$ band. Then, for each pixel, we compute its radius as
\[ a_i =\sqrt{\frac{[\Delta{x} \sin \theta+\Delta{y} \cos \theta]^2+[\Delta{x}\cos \theta+\Delta{y} \sin \theta]^2}{(1-\epsilon)^2}}, \]
where $\Delta{x}=(x_i-\bar x)$, $\Delta{y}=(y_i-\bar y)$, $\theta$ is the inclination angle and $\epsilon$ is the ellipticity.

We sorted these radii and binned them in 50 radial bins with approximately the same number of pixels, and computed for each of these bins the mean value of the H$\alpha$ emission. We then built the H$\alpha$ radial profile of each galaxy by plotting the mean flux in each bin as a function of the bin radius.

\begin{figure}
\includegraphics[width=1.0 \columnwidth,angle=0]{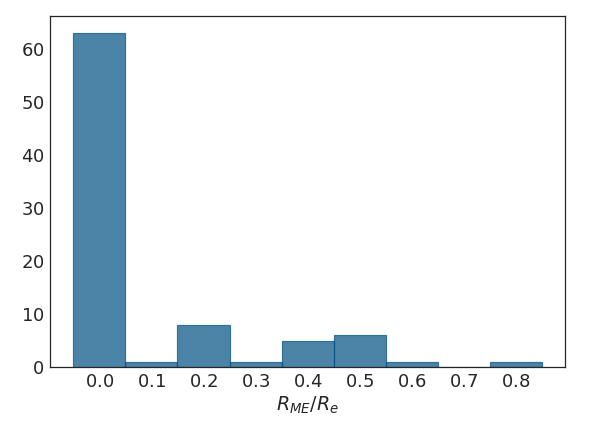}
\caption[]{The distribution of the radius of maximum H$\alpha$ emission, $R_{ME}$, for our sample. For almost 3/4 of our sample the maximum emission is in the galaxy center. }
\label{distribuicao_rmax}
\end{figure}

The resulting profiles are shown in Figures \ref{CE_profiles}, \ref{CL_profiles} and \ref{E_profiles} in the appendix \ref{profiles_ha_appendix}. From visual inspection, we first point out that they can be divided in two broad classes with respect to the maximum H$\alpha$ emission: those with maximum emission at the galaxy center, which we designate as $C$ (for central) profiles, and those where the peak of the emission is outside the galaxy center, which we designate as $EX$ (for extended) profiles. This is somewhat analogous to the two classes {\it centrally quiescent} and {\it centrally star-forming} of \cite{li2015} or the {\it unsuppressed} and {\it centrally suppressed} classes discussed by \cite{spindler2018}. Figure \ref{distribuicao_rmax} shows the distribution of the radius of maximum emission $R_{ME}$ for our sample. There are 63 and 23 galaxies of our sample in classes $C$ and $EX$, respectively.

\begin{figure*}
\begin{center}
\includegraphics[width=0.49 \columnwidth,angle=0]{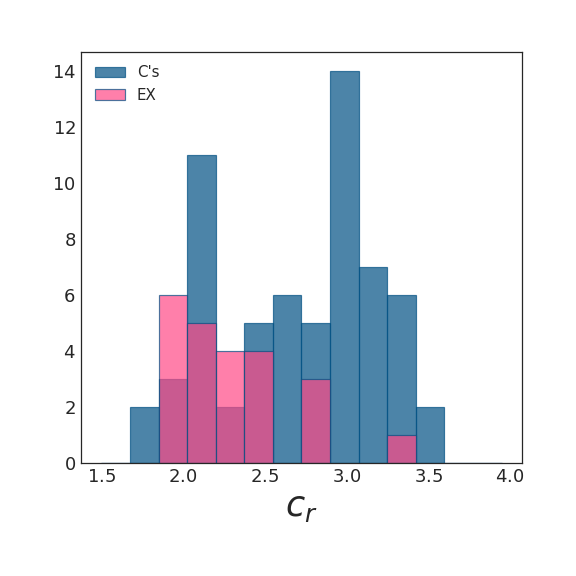}
\includegraphics[width=0.49 \columnwidth,angle=0]{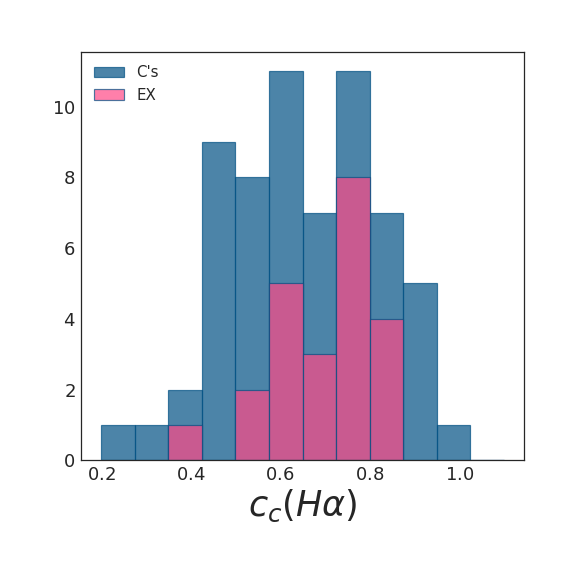}
\includegraphics[width=0.49 \columnwidth,angle=0]{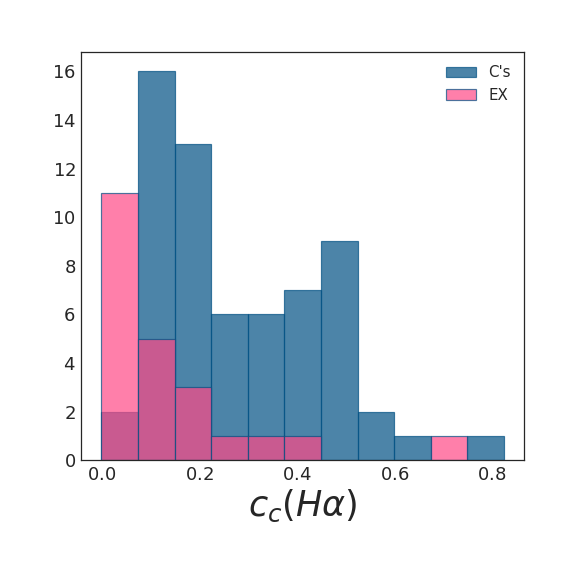}
\caption[]{Distribution of the concentrations $c_r$, $c_e(H\alpha)$ and $c_c(H\alpha)$ for the galaxies in the sample.}
\label{concentration_by_type}
\end{center}
\end{figure*}

A visual examination of $C$ profiles shows that they comprise a variety of behaviors, from those where most of the H$\alpha$ emission indeed comes from the center, to those where a significant fraction of this emission comes from more external regions of the galaxy. Figure \ref{concentration_by_type} shows the distribution of the concentrations $c_{e}(H\alpha)$, $c_{c}(H\alpha)$, and $c_r$ for the $C$ and $EX$ profile classes. From the examination of this and other figures, we decided to divide the $C$ profiles in two sub-classes, with respect to the $c_r$ concentration. Indeed, \cite{Strateva2001} has shown that this quantity is useful to discriminate between the red and blue galaxy populations, with  $c_r = 2.63$ providing a divisory line. We then define the $CE$ class as that containing early-type objects with $C$ profiles, i.e., those with $c_r > 2.63$, whereas the $CL$ class comprises late-type galaxies ($c_r < 2.63$) with $C$ profiles. From several tests, we verified that sub-classification of $C$ profiles with $c_r$ provides a more sensible classification than using the other concentrations, $c_{e}(H\alpha)$and $c_{c}(H\alpha)$. Our sample has 36 and 27 galaxies in classes $CE$ and $CL$, respectively.

\begin{figure*}
\includegraphics[width=0.85 \columnwidth,angle=0]{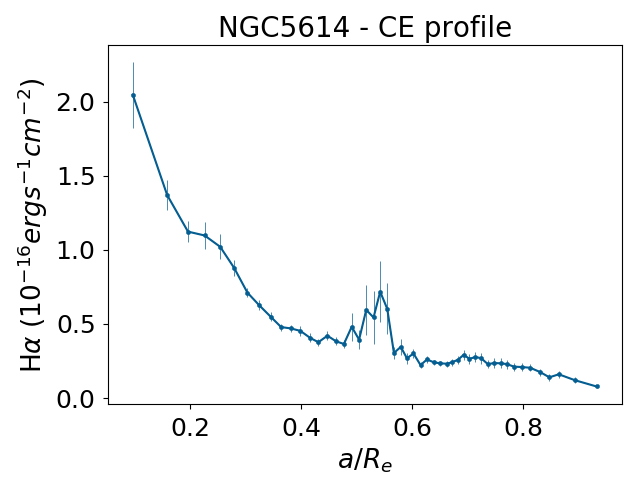}
\includegraphics[width=0.85 \columnwidth,angle=0]{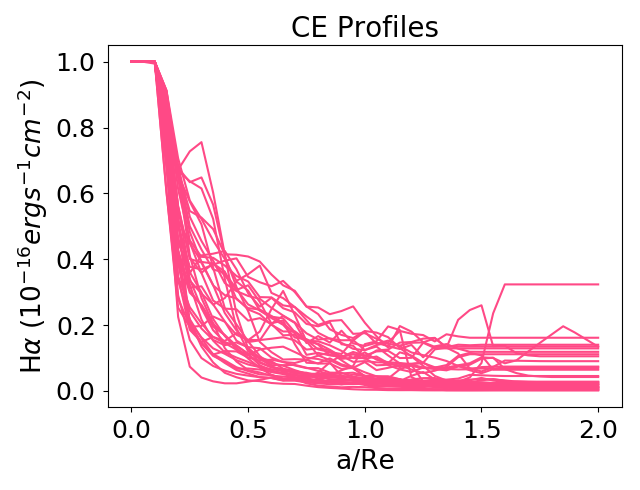}
\includegraphics[width=0.85 \columnwidth,angle=0]{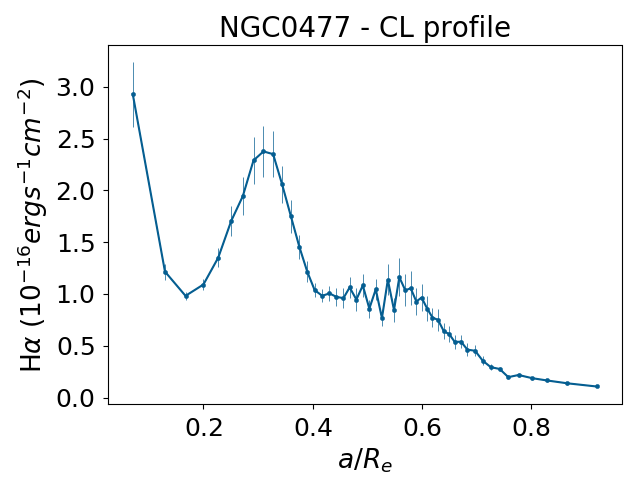}
\includegraphics[width=0.85 \columnwidth,angle=0]{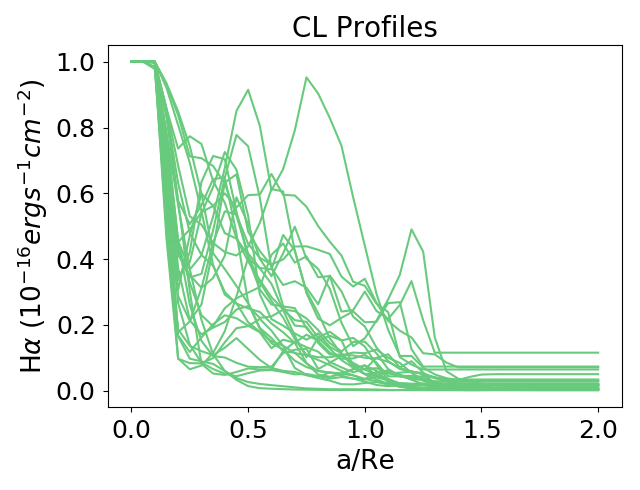}
\includegraphics[width=0.85 \columnwidth,angle=0]{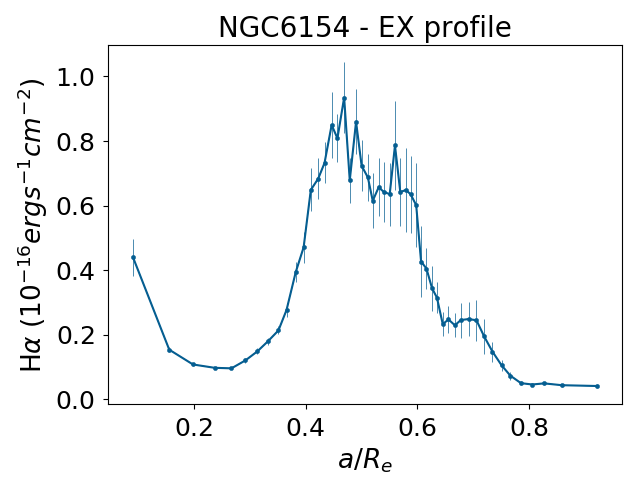}
\includegraphics[width=0.85 \columnwidth,angle=0]{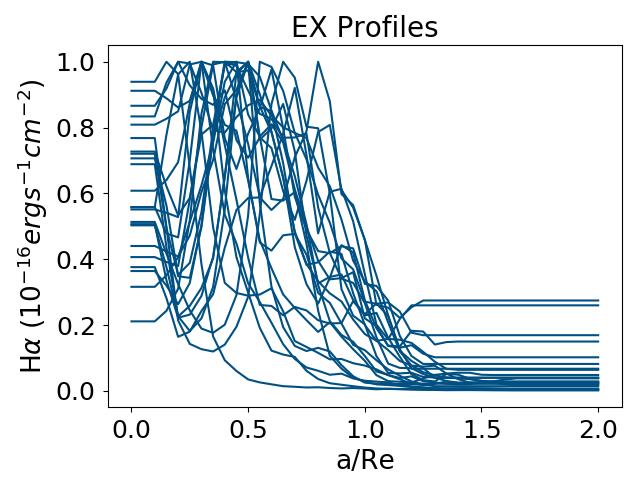}
\caption[]{Examples of the three H$\alpha$ profile types. From  top to bottom:  $CE$, $CL$ and $EX$ profiles. The left panels show a typical profile (that with the class median $c_r$), whereas the right panels shows all profiles of a given type together, normalized by their maximum H$\alpha$ emission.}
\label{perfil_ha_classificacao}
\end{figure*}

Figure  \ref{perfil_ha_classificacao} shows the result of this profile morphological analysis. The figure shows H$\alpha$ profiles separated by these three classes, as well as the profile  with median $c_r$ of each class. In general, $CE$ profiles present most of the emission comming from the galaxy center, without relevant emission outside the central region; $CL$ profiles tend to be more extended, indicating H$\alpha$ emission spread over larger radii. For type $EX$ profiles, by definition most of the emission comes from regions outside the galaxy center, in one or more bumps.


\begin{table}
\vspace{0.3cm}
\centering 
\begin{tabular}{|c| c| c| c|} 
\hline 
& $CE$ &  $CL$ &  $EX$ \\ 
\hline \hline                
E\underline{ }S0 & 28 & 1 &  1  \\
S\underline{ }early & 4 & 8 & 11  \\ 
S\underline{ }late & 4 & 18 & 11  \\ [1ex] 
\hline 
\end{tabular}
\caption[Tabela de contingência]{Contingency table for the H$\alpha$ profiles and morphological classes, where we show the number of galaxies in each profile and morphological class.} 
\label{Contigencytable_perfil_morfologia} 
\end{table}

\section{Results}
\label{sec:results}
In this Section we present some consequences of our profile classification, arguing that it is indeed meaningful, by examining relations between the profile types and other galaxy parameters: morphological classes, concentration measurements and  stellar population parameters. Finally, we use diagnostic diagrams to determine the origin of the H$\alpha$ emission.

\subsection{Relations with morphological types and concentrations}
\begin{figure}
\begin{center}
\includegraphics[width=0.98 \columnwidth,angle=0]{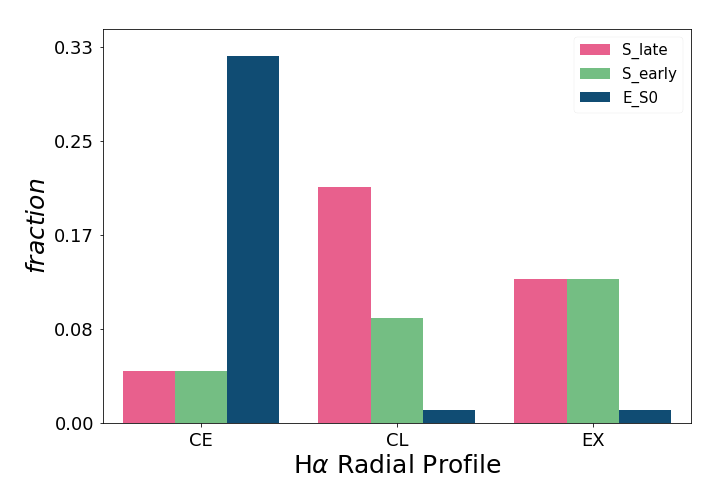}
\caption[]{Fraction of morphological classes in each of the H$\alpha$ profiles' type.}
\label{perfil_ha_mosaic}
\end{center}
\end{figure}

We start by investigating whether the profile types are related to the galaxy morphological classes. A simple $\chi^2$ test of the contingency table of morphological classes and profile types, Table \ref{Contigencytable_perfil_morfologia}, leads to $\chi^2 \simeq 53$, with a $p-value=8.5 \times 10^{-11}$, indicating that the profiles and the morphological classes are not independent. Figure \ref{perfil_ha_mosaic} shows the proportion of each morphological class for each profile type. Considering $C$ profiles, our two classes discriminate well between the E\underline{ }S0 (mostly in the $CE$ class) and S groups  (mostly in the $CL$ class). $EX$ profiles are mostly comprised of spiral galaxies. On the other side, the S\_early and S\_late groups are both numerous in $CL$ and $EX$ profile types. 

By analyzing the $CL$ profiles in Figure \ref{CL_profiles} individually, it may be noticed that $\sim75\%$ of the objects have at least a second H$\alpha$ emission peak along the galaxy's extension, probably coming from star formation in the disk of these objects. Since $CL$ profiles are dominated by spiral galaxies, both S\underline{ }early and S\underline{ }late, it is reasonable to associate these secondary H$\alpha$ emission peaks to star-forming arms or rings in the body of the galaxy.
In contrast, when we examine E\underline{ }S0 objects classified as CL and EX (NGC1349 and NGC5784, respectively), we notice that they also have emission peaks outside the center. As \cite{gomes2016a} argues, these early-type galaxies present evidence of  recent stellar activity, showing that not all early-type galaxies are actually quenched.

\begin{figure*}
\begin{center}
\includegraphics[width=0.65 \columnwidth,angle=0]{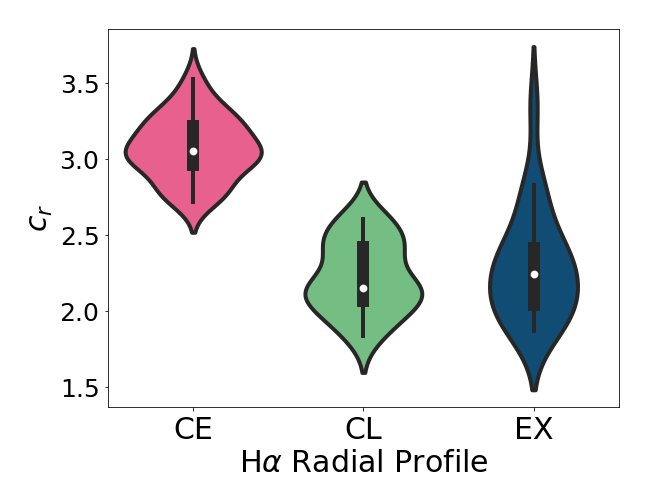}
\includegraphics[width=0.65 \columnwidth,angle=0]{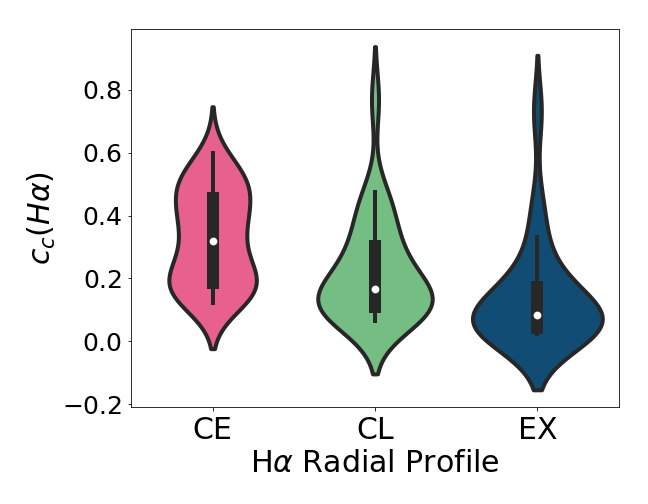}
\includegraphics[width=0.65 \columnwidth,angle=0]{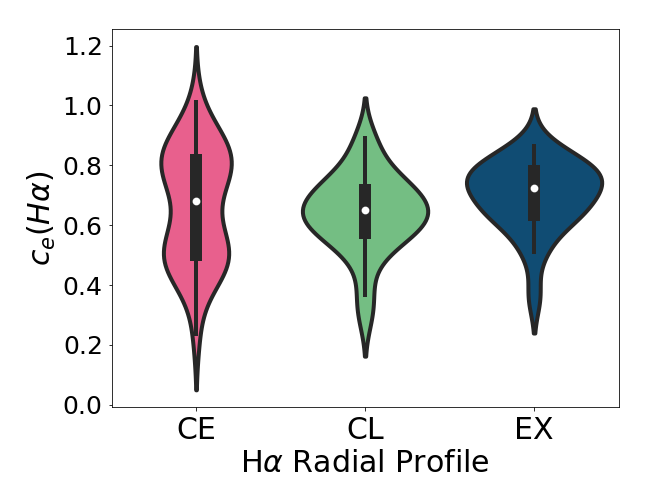}
\caption[]{Distribution of concentrations $c_r$, $c_c(H\alpha)$ and $c_e(H\alpha)$ with their mean value shown as a white dot inside the violin plot.}
\label{fig: concentracoes_violin}
\end{center}
\end{figure*}

The relation between concentrations and the three types of H$\alpha$  profiles is shown in Figure \ref{fig: concentracoes_violin} using violin diagrams (which depict the range of values and the probability of variables). The mean value of the concentration $c_r$ is larger for type $CE$ than for $CL$, by construction, and  $CL$ and $EX$ types have similar $c_r$ values, corresponding approximately to the blue peak seen in the bimodal $c_r$ distribution in Figure \ref{figure: distribuicoes_concentracao}.  The mean value of the central H$\alpha$ concentration decreases from $CE$ to $CL$ and to $EX$ profiles. The effective concentration, which is in fact measuring how extended is the line emission in the galaxy, increases from $CE$ to  $CL$ to $EX$. Table \ref{tabela_propriedades_Estelares_perfil} presents the mean values of the concentration measurements for each profile type.

\subsection{Stellar Population Properties}

\begin{figure*}
\begin{center}
\includegraphics[width=0.65 \columnwidth,angle=0]{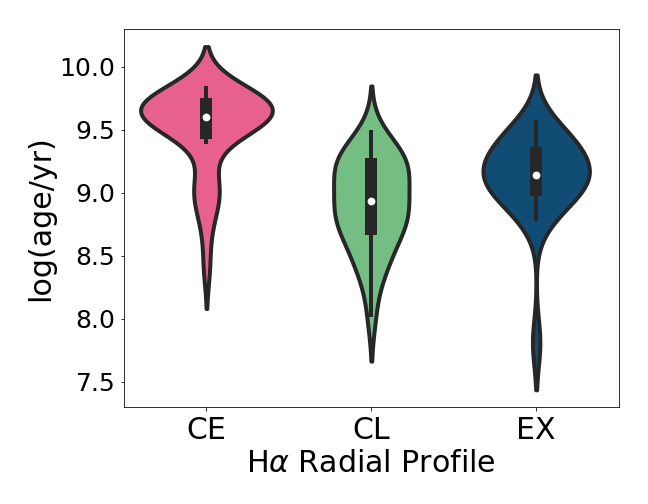}
\includegraphics[width=0.65 \columnwidth,angle=0]{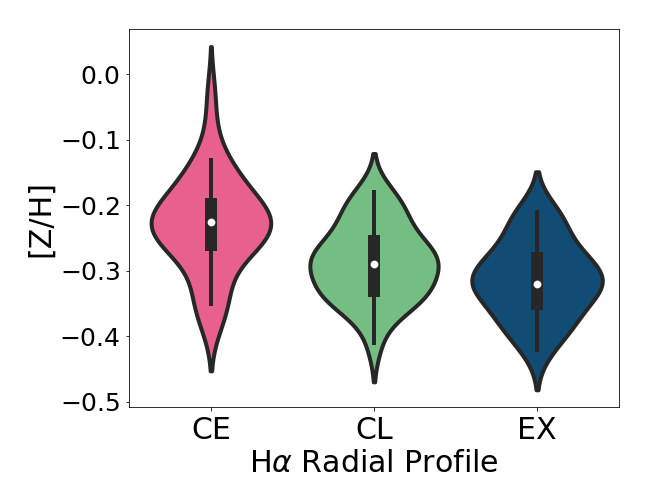}
\includegraphics[width=0.65 \columnwidth,angle=0]{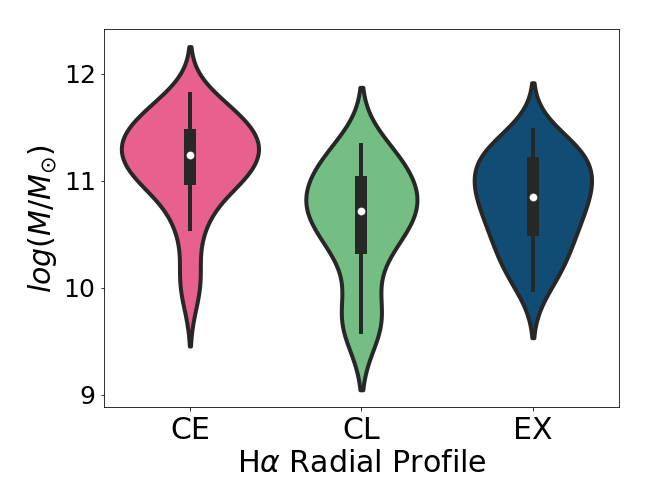}
\caption[]{The violin plot for the stellar age, metallicity and stellar mass and for each type of H$\alpha$ profile.}
\label{perfis_stellar_properties}
\end{center}
\end{figure*}

To investigate the relations of the H$\alpha$ profiles with the properties of stellar populations, we have also used the mean values of some stellar population properties- age, metallicity and stellar mass- obtained from Pipe3d, since these quantities are good tracers of other properties of galaxies \citep[][for some examples]{gallazzi2005,gallazzi2008,peletier2013,rosa2014b,zibetti2017}. The mean values for these quantities for each type of profile are shown in Table \ref{tabela_propriedades_Estelares_perfil}, whereas their violin diagrams, by H$\alpha$ profile types, are shown in Figure \ref{perfis_stellar_properties}.

\begin{table*}
\vspace{0.3cm}
\centering 
\begin{tabular}{|c| c| c| c| c| c| c|} 
\hline 
Profile & ${c_r}$ & ${c_c}(H\alpha)$ & ${c_e}(H\alpha)$ & log($t^*$/Gyr)  & log($Z/Z_{\odot}$) & log($M_* / M_{\odot}$)\\ 
\hline \hline
CE & 3.07$\pm 0.03$ & 0.33$\pm0.02$ & 0.66$\pm 0.03$	& 9.49$\pm0.05$	& -0.23$\pm 0.01$ &	11.15$\pm0.07$ \\			
CL & 2.22$\pm 0.04$ & 0.22$\pm0.03$ & 0.64$\pm 0.02$	& 8.90$\pm0.07$	& -0.29$\pm 0.01$ &	10.61$\pm0.10$ \\			
EX  & 2.29$\pm 0.07$ & 0.14$\pm0.03$ & 0.70$\pm 0.02$	& 9.12$\pm0.07$	& -0.31$\pm 0.01$ &	10.80$\pm0.09$ \\	[1ex] 
\hline 
\end{tabular}
\caption[]{Mean value and standard deviation for the concentrations and stellar population parameters for the three  types of H$\alpha$ profiles.} 
\label{tabela_propriedades_Estelares_perfil} 
\end{table*}

Considering only $C$ types, the mean ages, metallicities and masses are larger for galaxies with $CE$ than with $CL$ profiles; comparing $CL$ and $EX$ types, we notice that the latter seems to represent a population less evolved than the former, in the sense that their mean ages and metallicities are smaller for  $CL$ than for $EX$ classes. Stellar masses in our sample are larger for  $CE$ than for $CL$ types, and are larger for $EX$ than $CL$ types.

\subsection{The source of H$\alpha$ emission}
\label{emission_lines_source}

\begin{figure*}
\begin{center}
\includegraphics[width=0.85 \columnwidth,angle=0]{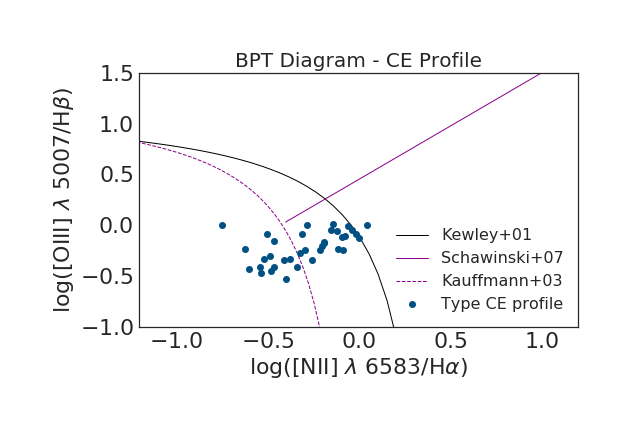}
\includegraphics[width=0.85 \columnwidth,angle=0]{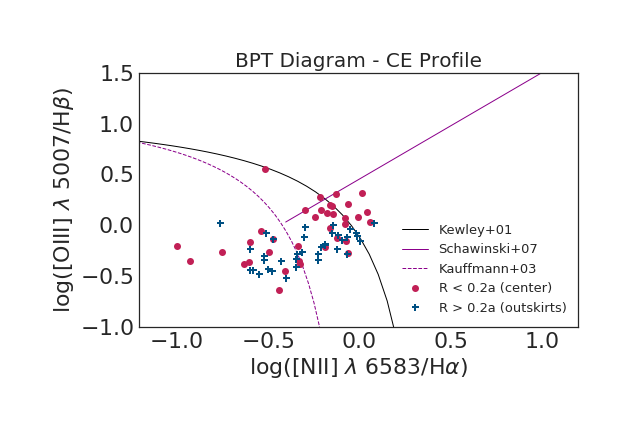}
\includegraphics[width=0.85 \columnwidth,angle=0]{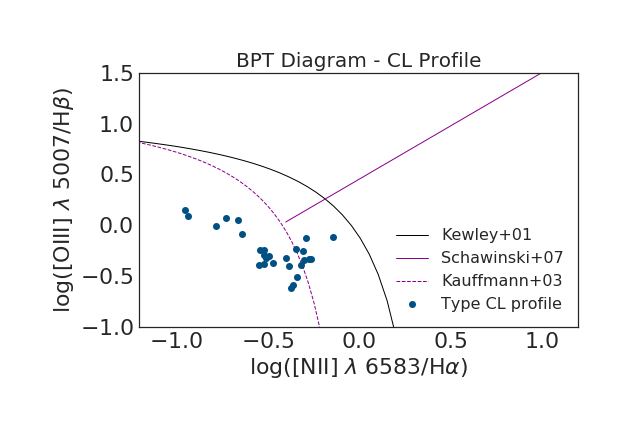}
\includegraphics[width=0.85 \columnwidth,angle=0]{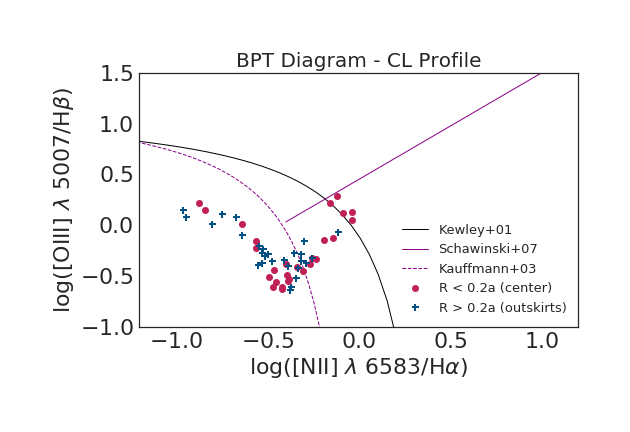}
\includegraphics[width=0.85 \columnwidth,angle=0]{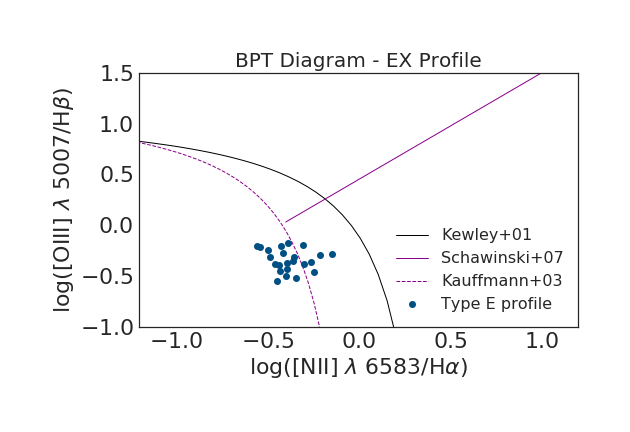}
\includegraphics[width=0.85 \columnwidth,angle=0]{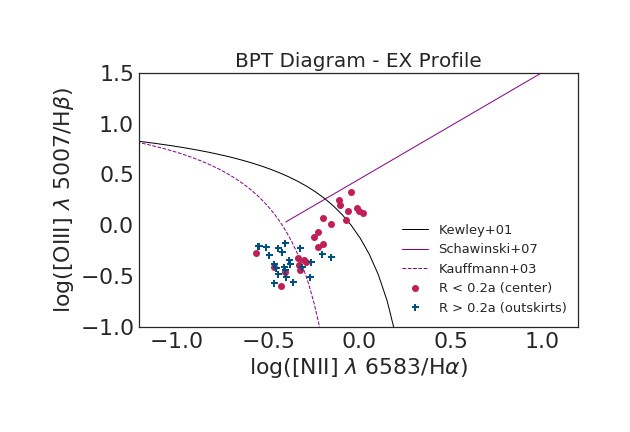}
\caption[]{BPT diagram for galaxies in the sample: left panels show the diagram for galaxy mean emission line ratios of each type of H$\alpha$ profile. Right panels present the BPT diagram for the different profile types separated by  inner (in blue) and outer (in pink) regions, respectively.}
\label{bpt_all_galaxies}
\end{center}
\end{figure*}

The next step in our analysis is to investigate the nature of the H$\alpha$ emission associated with each type of profile, if star formation, some kind of nuclear activity or both. For this we have made use of the \cite{bpt1981} BPT diagram. Figure \ref{bpt_all_galaxies} presents BPT diagrams for each profile type. At the left side of this figure we show the mean BPT diagram  for each galaxy of a given profile type, whereas in the right panels we show this diagram separately for the internal ($r < 0.2 a$, in red) and external ($r > 0.2a$, in blue) regions of galaxies, again separated by profile type. This figure also shows the empirical line proposed by \cite{kauffmann2003} (where galaxies below it are classified as pure star-forming objects), the model line proposed by \cite{kewley2001} (where galaxies above it are considered AGNs), and the line proposed by \cite{schawinski2007} to discriminate between LINERs (below the line) and Seyfert galaxies (above the line). Any object between Kauffmann's line and Kewley's line are considered as an object in transition, with emission characteristics of both stars and AGNs.

The BPT diagram for the galaxies as a whole shows interesting trends with the H$\alpha$ profile. $CE$ objects present emission coming from both the star-forming and transition regions, with just one galaxy in the LINER region of the diagram. The fraction of galaxies in the transition region decreases from type $CE$ to $EX$ to $CL$ and, at the same time, the proportion of objects in the star-forming region increases. Galaxies with H$\alpha$ profiles of type $EX$ in this panel are all close to the line proposed by \cite{kauffmann2003}, showing less scatter than the other profiles.

Panels at the right side of Figure \ref{bpt_all_galaxies} tell a more interesting story. Considering the central region of the galaxies ($r < 0.2 a$, in red), type $CE$ profiles have most of their galaxies, $\sim 47\%$, in the transition region, $\sim 36 \%$ in the star-formation region and the other $17\%$ coming from the LINER region of the BPT diagram. Profiles of type $CL$ have $63\%$, $22\%$ and $15\%$ of central region emission coming from star-forming, transition and LINER regions, respectively. Galaxies with $CL$ profiles in the star-forming region are distributed forming a tail towads low values of $[NII]/H\alpha$, suggesting that these objects have lower metallicity than the others in $CE$ and $EX$ classes. $EX$ galaxies present a larger homogeneity in the distribution of central emission, with $30\%$, $39\%$ and $30\%$ of their members in star-forming, transition and LINER regions, respectively.

\section{Discussion}
\label{sec:discussion}

The H$\alpha$ emission is by far one of the most important tools to investigate galaxy properties. In this paper, through an analysis of a sample of CALIFA galaxies, we present a classification of the H$\alpha$ radial profile in three classes: $CE$, $CL$ and $EX$. The central profiles $CE$ and $CL$, have the peak of the line emission in the galaxy centre, whereas in $EX$ profiles, the maximum H$\alpha$ emission is outside the center. We found useful to distinguish two classes of central profiles ($CE$ and $CL$), dominated by early- or late-type populations, accordingly with the value of the $c_r$ concentration \citep{Strateva2001}. The profiles' types exhibit good relation with some structural and physical parameters such as concentrations and stellar population properties.

The first aspect to point out is that the large majority of our sample, $\sim 75$\%, has central ($C$) profiles. Dividing this group by the $c_r$ value in early, $CE$, and late $CL$, subtypes, we noticed that  $CE$ profiles are dominated by early-type galaxies, E and S0. Spiral galaxies in the $CL$ class are mostly blue objects (in terms of the $c_r$ criterion), while the $EX$ class contains the same fraction of blue and red late-type galaxies. In other words, most of the spirals with central profiles are blue galaxies, whereas those with an extended profile have a mix of red and blue objects. E\underline{ }S0 galaxies are less than 4\% of the $CL$ and $EX$ profile classes; early-type galaxies in these classes are probably objects still forming stars,  not quenched yet \citep[e.g.][]{gomes2016a}.

Concerning the  H$\alpha$ concentration parameters, we have shown that the central concentration $c_c$ is indeed useful to describe the emission profiles: for central profiles, $CE$ objects tend to have more H$\alpha$ emission in the central region than the $CL$s and $EX$s. The effective concentration does not show strong trends with the profile types, although it tends to be larger for $EX$ profiles than for the others. 

Stellar population properties also show trends with the profile types. Stellar populations are older for $CE$ profiles. Comparing mean stellar ages for  $CL$ to $EX$ profiles, we can verify that the latter presents older populations than the former. Mean metallicities  decrease from $CE$ to $CL$ and to $EX$. These results indicate that $CE$ profiles are more likely to be associated to early-type objects than $CE$ and $EX$ profiles.

The BPT diagram also presents significant differences between the three classes. Considering the mean value of the line ratios (left panels of Fig. \ref{bpt_all_galaxies}), the most striking difference is the large fraction of transition objects for $CE$ profiles compared to the others, which have most of their galaxies in the star-forming part of the diagrams. When we consider the internal and external BPT diagrams, several interesting features appear. For all profiles the number of galaxies with LINER central emission increases. The same is true for objects in the transition region, at least for late-types. For these galaxies, the large majority of the emission in the external regions seems to be due to star formation.

Indeed, the second aspect we consider important to highlight here is the difference in the properties of late type galaxies in the $CL$ and $EX$ classes. $EX$ objects have a significant fraction with central emission in the transition region, whereas for $CL$ objects, the fraction where the emission can be ascribed to star-formation is larger. Also, $CL$ galaxies  seem to have a larger range of physical properties (e.g. nebular abundances) than late types in the $EX$ class. This seems to indicate that $EX$ objects are more evolved than $CL$ galaxies.

Our results support the findings of \cite{li2015} and \citet{spindler2018}. The latter authors analyzed SFR profiles from the SDSS-IV MaNGA survey, showing that they can be divided in 'centrally suppressed' and 'unsuppressed' star formation, in close association with our C and EX profiles. Our conclusions are in general agreement with theirs; they found that both centrally suppressed and unsuppressed galaxies have a bimodal distribution, whereas our EX galaxies are mostly blue. This is probably a consequence of our small sample size.

The classification procedure adopted here is objective and can be extended to other IFU-like observational data. We are interested in using this approach to investigate galaxy samples  in the local universe observed by the surveys S-PLUS \citep[][in prep.]{mendesdeoliveira2018}, J-PLUS \citep{cenarro2018}, and J-PAS \citep{benitez2014}. Despite being photometric surveys, they are imaging large areas of the sky with a large number of narrow and broad-band filters (12 for S-PLUS and J-PLUS, and 59 for J-PAS). They contain narrow filters centred on H$\alpha$ and it has been shown that the H$\alpha$ emission can indeed be reliably determined for nearby objects with the observational setup of these surveys \citep{vilella-rojo2015,logrono-garcia2018}.


\section{Summary}
\label{sec:summary}

We have determined the radial profile of the  H$\alpha$ emission for a sample of 86 face-on galaxies from the CALIFA survey. After visual examination, we classified these profiles in three classes, $CE$, $CL$ and $EX$, taking in to account the position of the maximum H$\alpha$ emission (in the galaxy centre or not) and, for galaxies where the maximum of H$\alpha$ emission comes from the galaxy centre, if the dominant population is red or blue, by using the light concentration in the $r$-band, $c_r$. Our results show that, for $\sim 75$\% of the sample, the peak of the emission is in the galaxy centre.

Galaxy properties, like morphology, mean stellar ages and metallicities, and the nature of the line emission, correlate well with these three classes. For example, most of the objects in the $CE$ class are early-type galaxies. Late-type galaxies are spread between the $CL$  and $EX$ classes, with  $CL$ objects presenting features of a less evolved population when compared to the $EX$ class.

The classification presented here helps to highlight the diversity of  H$\alpha$ emission in the galaxies of the local universe and its close links with properties usually adopted to describe the galaxy population.

\section*{Acknowledgements}

PMN thanks the financial support from Brazilian funding agency CNPq (grant 142436/2014-3). She also thanks Rosa González Delgado, Enrique Perez and Ruben García-Benito for useful comments. PMN also thanks the Instituto de Astrof\'isica de Andaluc\'ia (IAA/CSIC) and the Instituto de Astronomia, Geofísica e Ciências Atmosféricas (IAG/USP) for their warm scientific environment that made this work possible. LSJ aknowledges support of CNPq (304819/2017-4) and FAPESP (2017/237666-0) to his work.




\bibliographystyle{mnras}
\bibliography{bibliografia} 

\appendix

\section{H$\alpha$ Profiles}

\label{profiles_ha_appendix}
\begin{figure*}
\begin{center}
\includegraphics[width=0.411\columnwidth,angle=0]{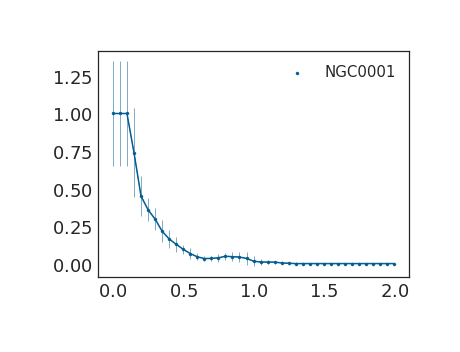}
\includegraphics[width=0.411\columnwidth,angle=0]{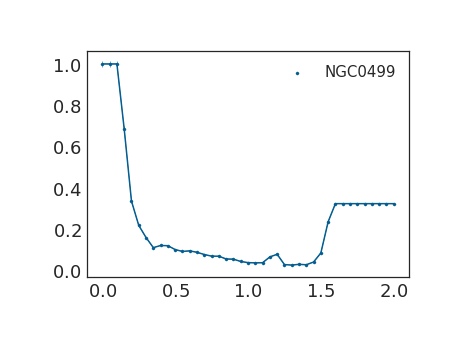}
\includegraphics[width=0.411\columnwidth,angle=0]{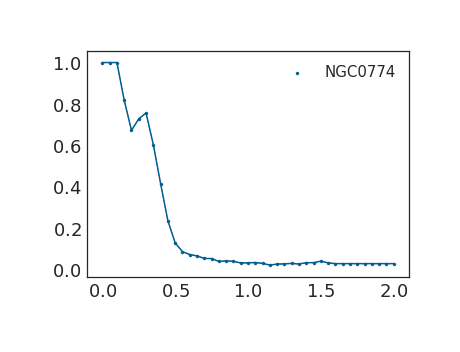}
\includegraphics[width=0.411\columnwidth,angle=0]{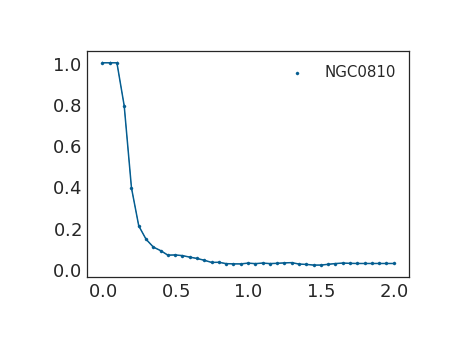}
\includegraphics[width=0.411\columnwidth,angle=0]{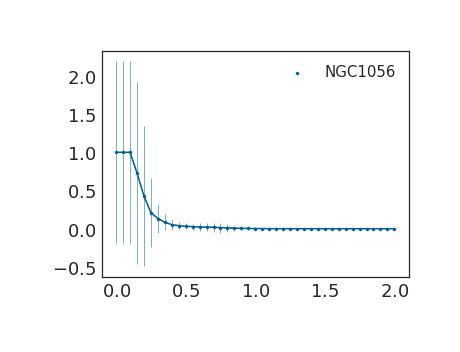}
\includegraphics[width=0.411\columnwidth,angle=0]{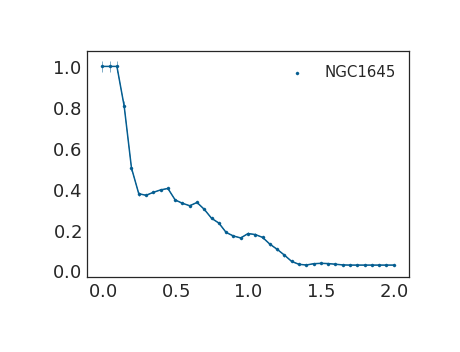}
\includegraphics[width=0.411\columnwidth,angle=0]{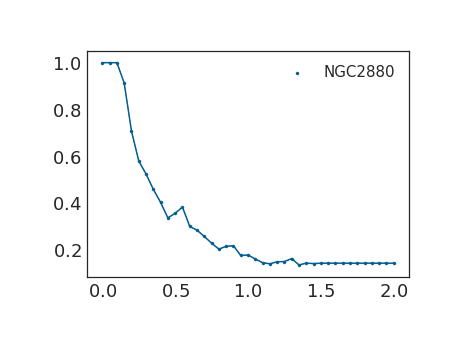}
\includegraphics[width=0.411\columnwidth,angle=0]{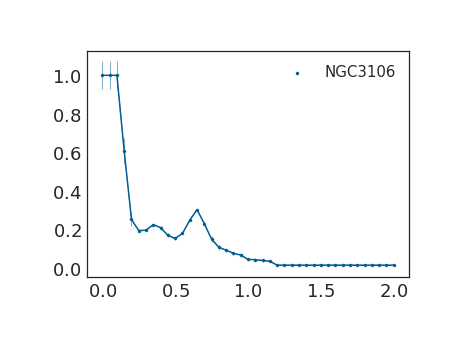}
\includegraphics[width=0.411\columnwidth,angle=0]{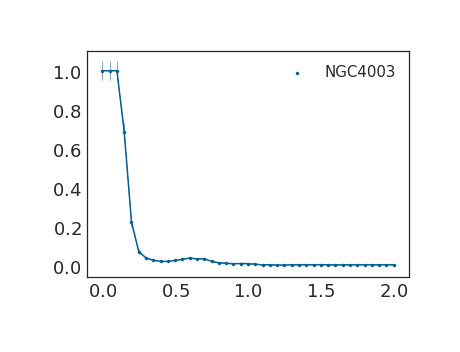}
\includegraphics[width=0.411\columnwidth,angle=0]{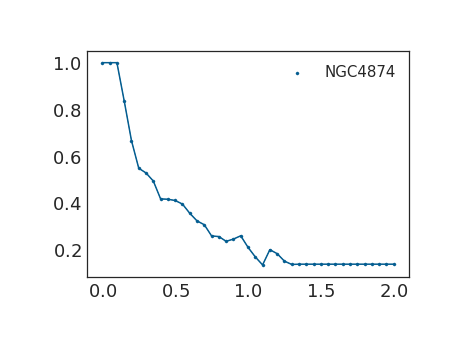}
\includegraphics[width=0.411\columnwidth,angle=0]{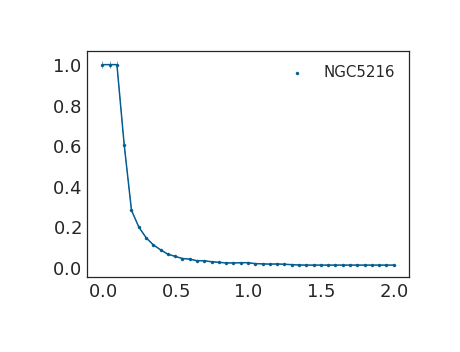}
\includegraphics[width=0.411\columnwidth,angle=0]{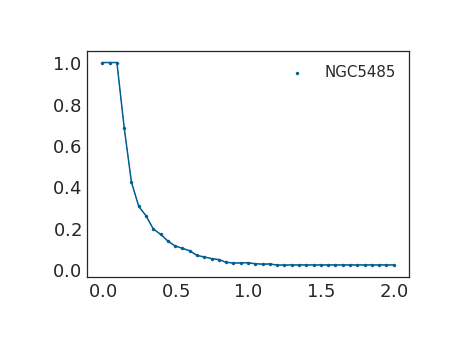}
\includegraphics[width=0.411\columnwidth,angle=0]{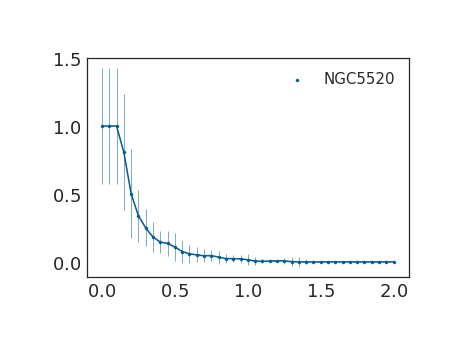}
\includegraphics[width=0.411\columnwidth,angle=0]{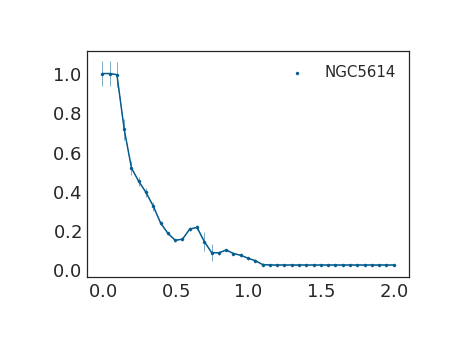}
\includegraphics[width=0.411\columnwidth,angle=0]{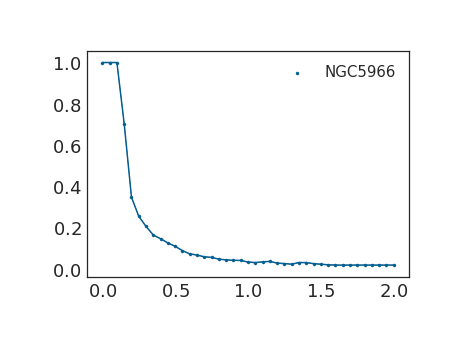}
\includegraphics[width=0.411\columnwidth,angle=0]{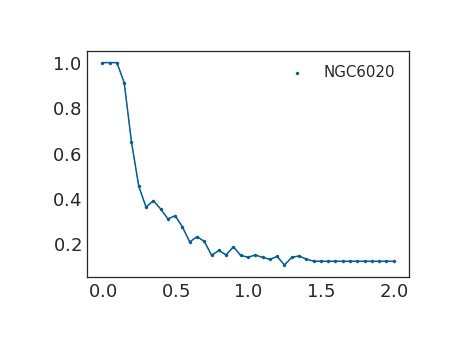}
\includegraphics[width=0.411\columnwidth,angle=0]{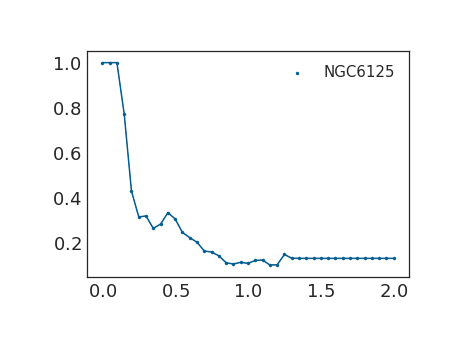}
\includegraphics[width=0.411\columnwidth,angle=0]{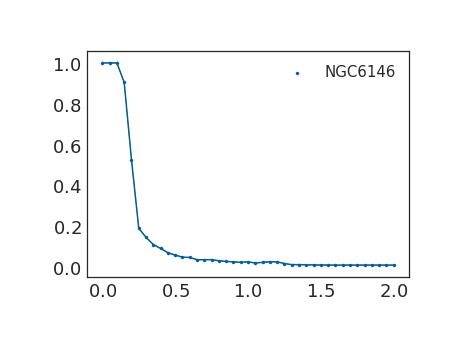}
\includegraphics[width=0.411\columnwidth,angle=0]{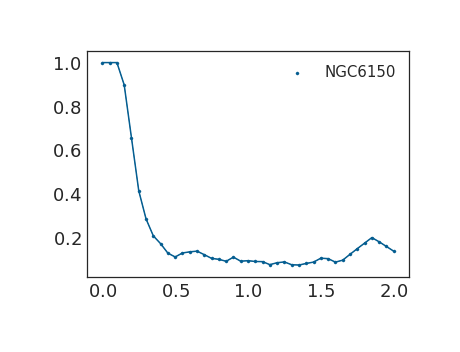}
\includegraphics[width=0.411\columnwidth,angle=0]{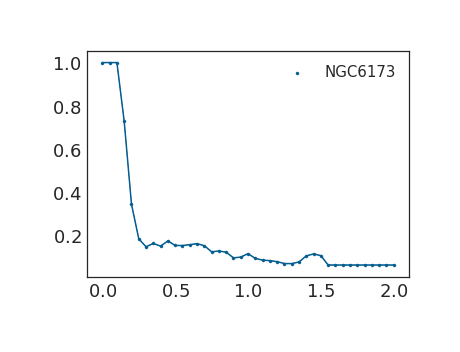}
\includegraphics[width=0.411\columnwidth,angle=0]{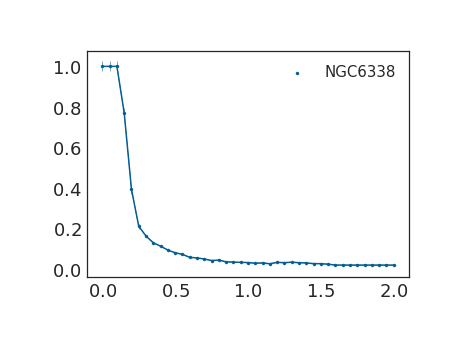}
\includegraphics[width=0.411\columnwidth,angle=0]{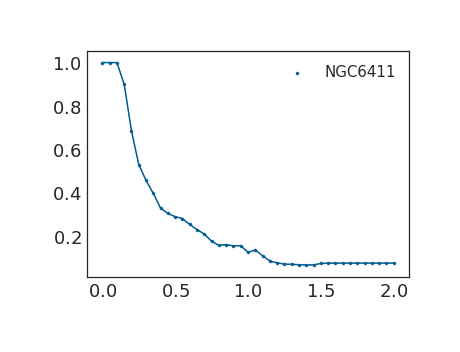}
\includegraphics[width=0.411\columnwidth,angle=0]{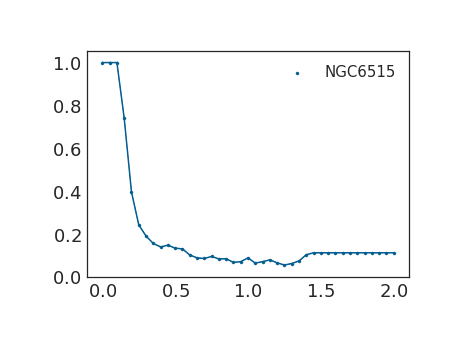}
\includegraphics[width=0.411\columnwidth,angle=0]{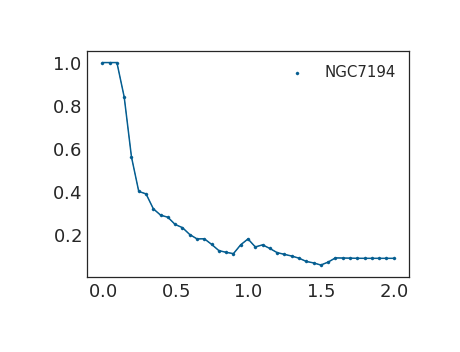}
\includegraphics[width=0.411\columnwidth,angle=0]{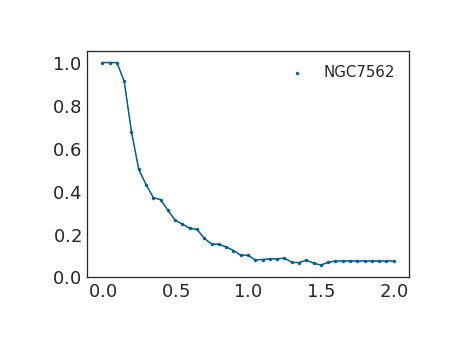}
\includegraphics[width=0.411\columnwidth,angle=0]{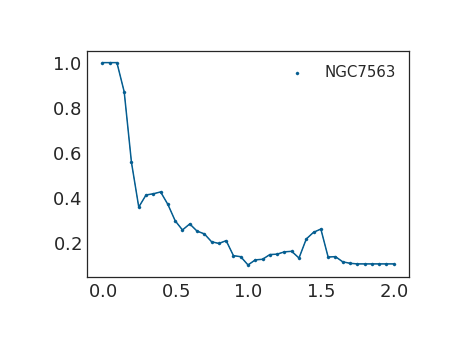}
\includegraphics[width=0.411\columnwidth,angle=0]{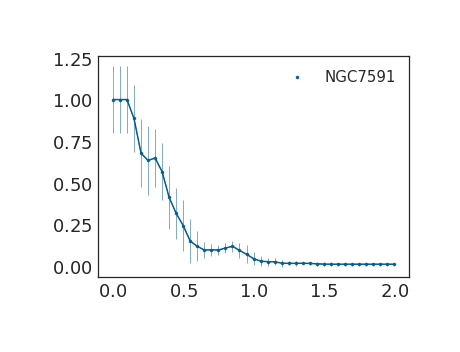}
\includegraphics[width=0.411\columnwidth,angle=0]{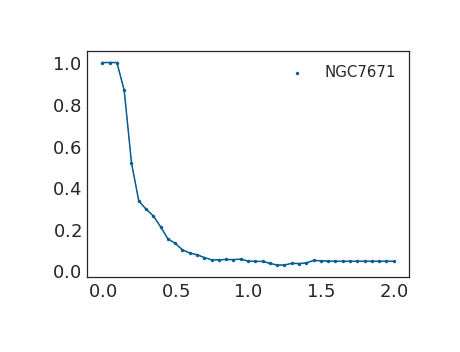}
\includegraphics[width=0.411\columnwidth,angle=0]{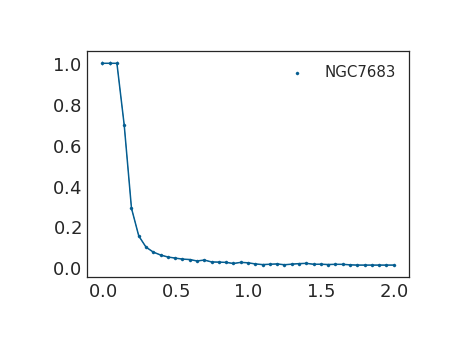}
\includegraphics[width=0.411\columnwidth,angle=0]{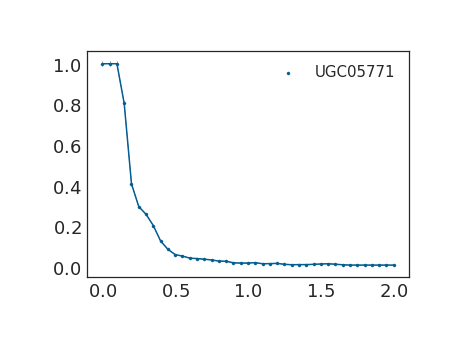}
\includegraphics[width=0.411\columnwidth,angle=0]{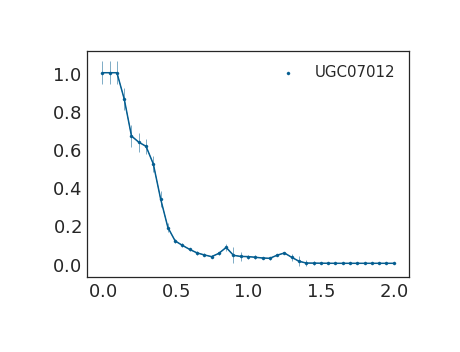}
\includegraphics[width=0.411\columnwidth,angle=0]{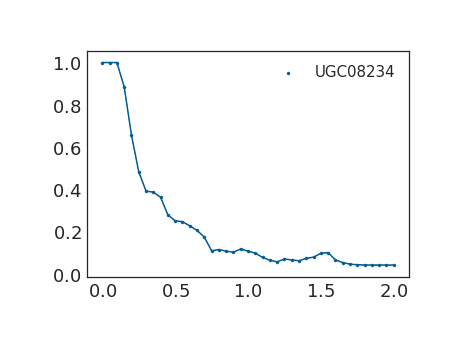}
\includegraphics[width=0.411\columnwidth,angle=0]{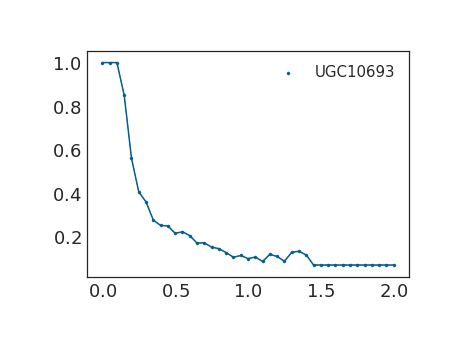}
\includegraphics[width=0.411\columnwidth,angle=0]{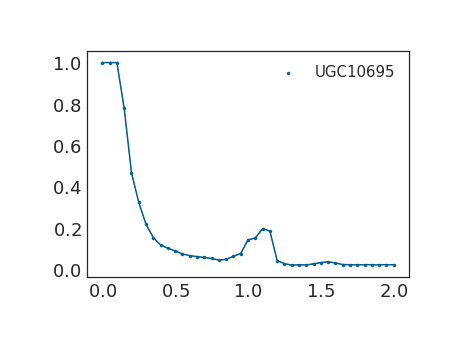}
\includegraphics[width=0.411\columnwidth,angle=0]{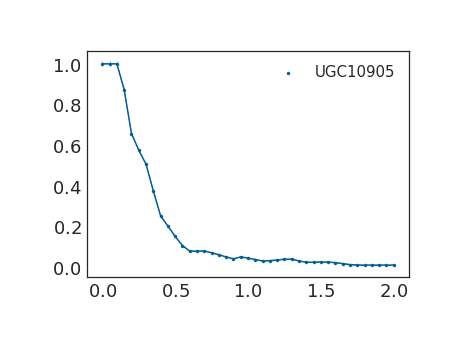}
\includegraphics[width=0.411\columnwidth,angle=0]{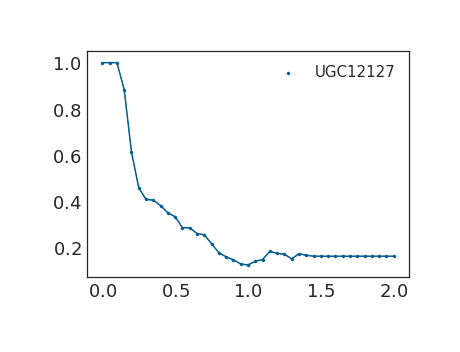}
\caption[]{CE Profiles, in units of $10^{-16} erg s^{-1} cm^{-2}$ for the H$\alpha$ emission.}
\label{CE_profiles}
\end{center}
\end{figure*}

\begin{figure*}
\begin{center}
\includegraphics[width=0.411\columnwidth,angle=0]{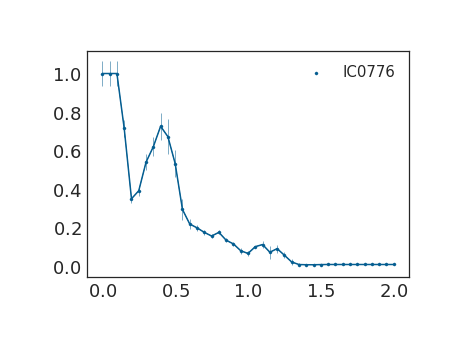}
\includegraphics[width=0.411\columnwidth,angle=0]{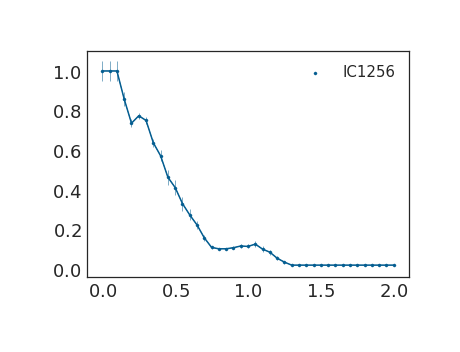}
\includegraphics[width=0.411\columnwidth,angle=0]{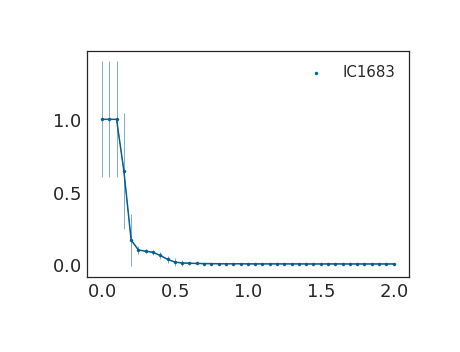}
\includegraphics[width=0.411\columnwidth,angle=0]{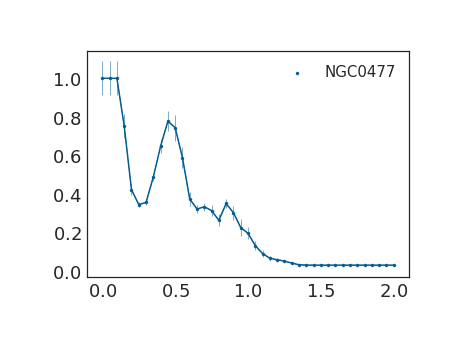}
\includegraphics[width=0.411\columnwidth,angle=0]{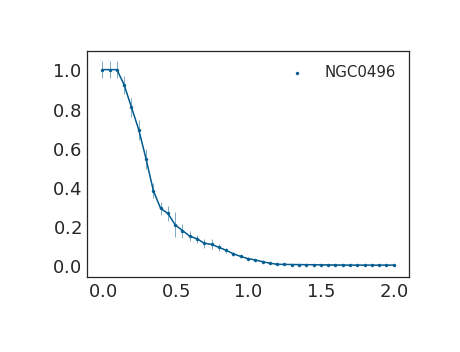}
\includegraphics[width=0.411\columnwidth,angle=0]{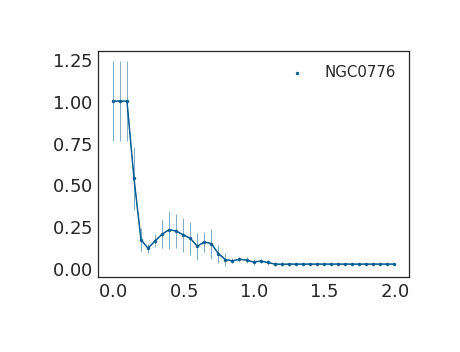}
\includegraphics[width=0.411\columnwidth,angle=0]{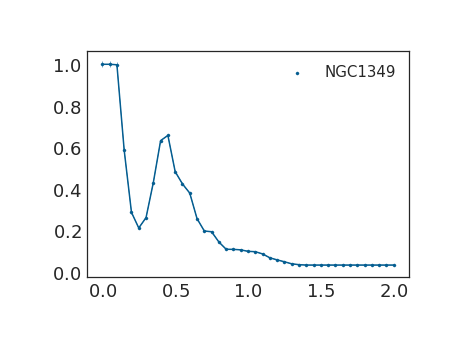}
\includegraphics[width=0.411\columnwidth,angle=0]{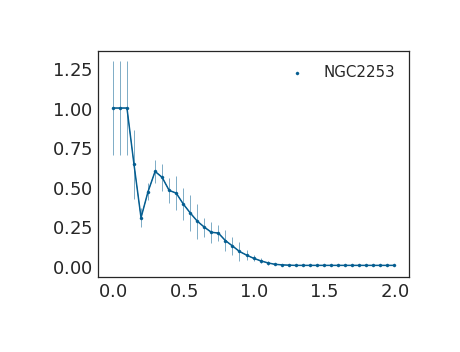}
\includegraphics[width=0.411\columnwidth,angle=0]{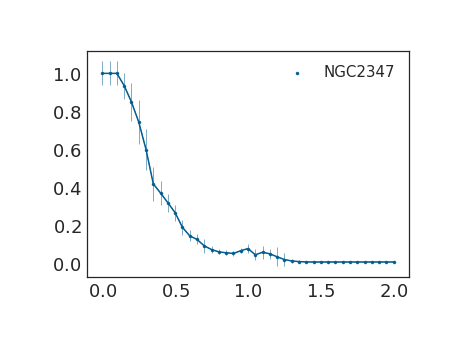}
\includegraphics[width=0.411\columnwidth,angle=0]{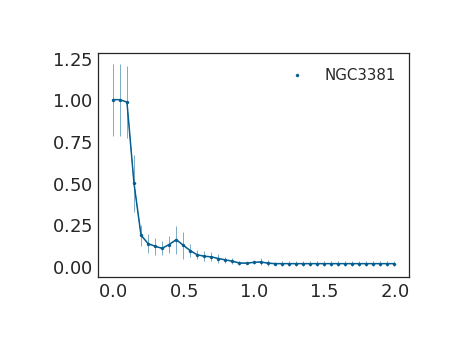}
\includegraphics[width=0.411\columnwidth,angle=0]{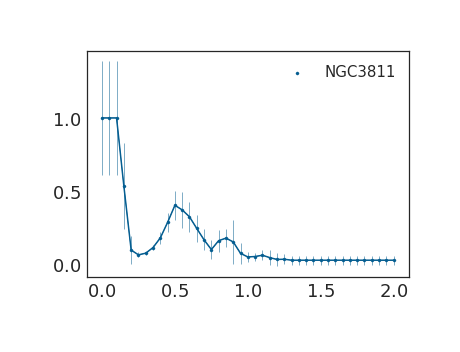}
\includegraphics[width=0.411\columnwidth,angle=0]{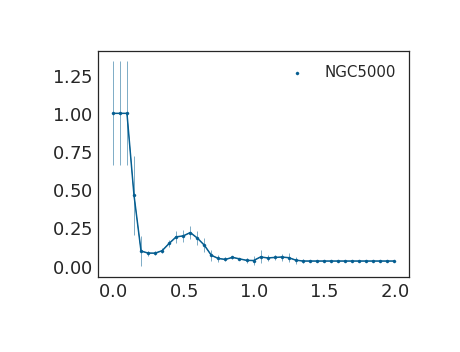}
\includegraphics[width=0.411\columnwidth,angle=0]{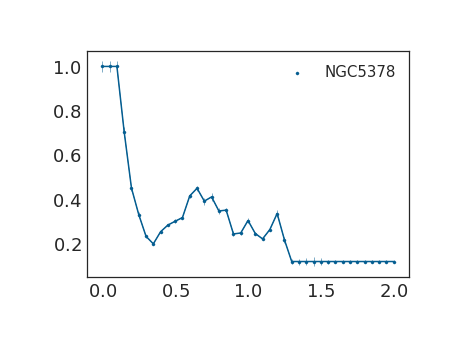}
\includegraphics[width=0.411\columnwidth,angle=0]{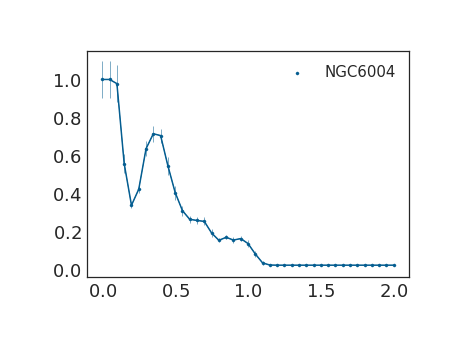}
\includegraphics[width=0.411\columnwidth,angle=0]{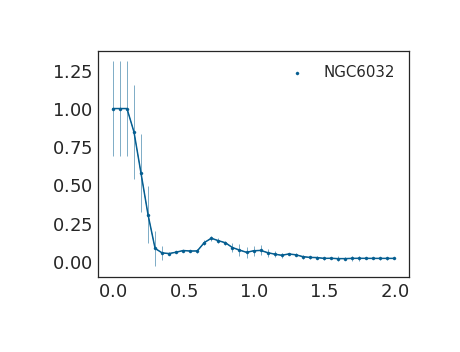}
\includegraphics[width=0.411\columnwidth,angle=0]{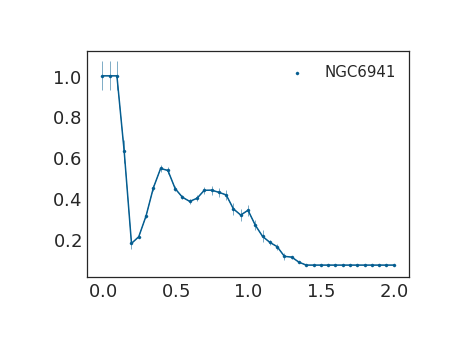}
\includegraphics[width=0.411\columnwidth,angle=0]{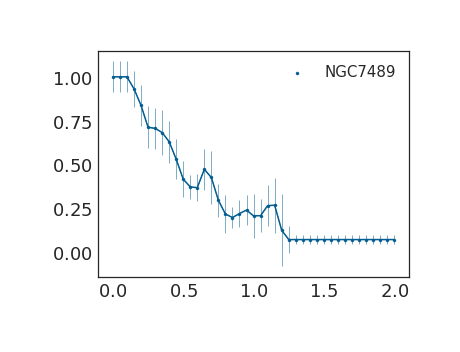}
\includegraphics[width=0.411\columnwidth,angle=0]{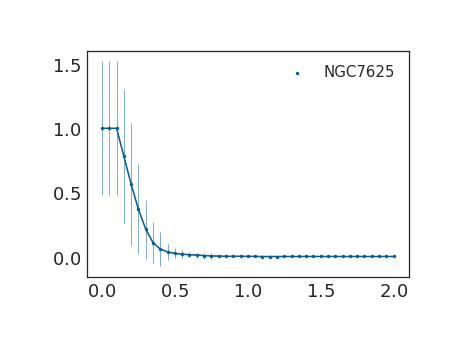}
\includegraphics[width=0.411\columnwidth,angle=0]{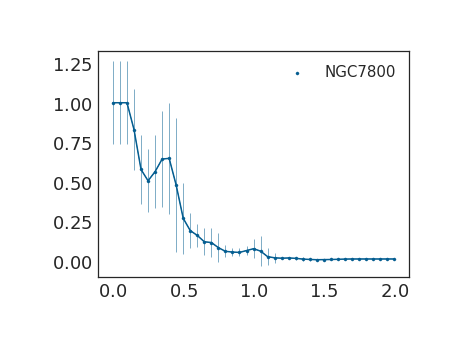}
\includegraphics[width=0.411\columnwidth,angle=0]{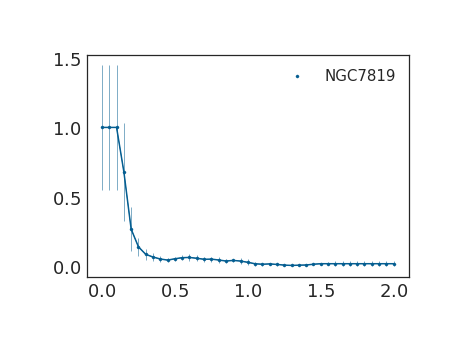}
\includegraphics[width=0.411\columnwidth,angle=0]{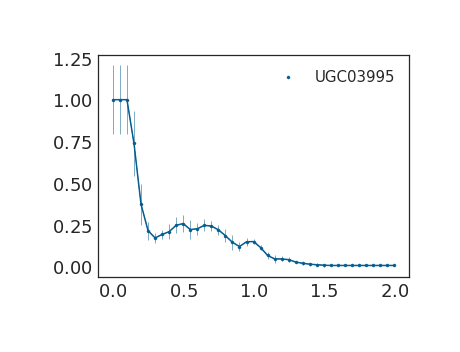}
\includegraphics[width=0.411\columnwidth,angle=0]{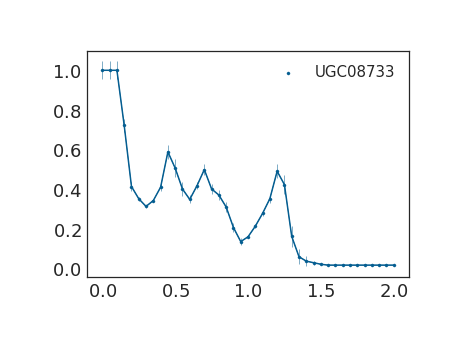}
\includegraphics[width=0.411\columnwidth,angle=0]{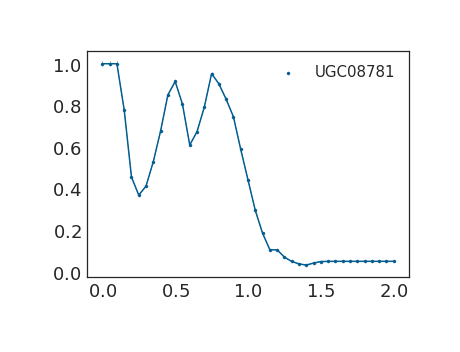}
\includegraphics[width=0.411\columnwidth,angle=0]{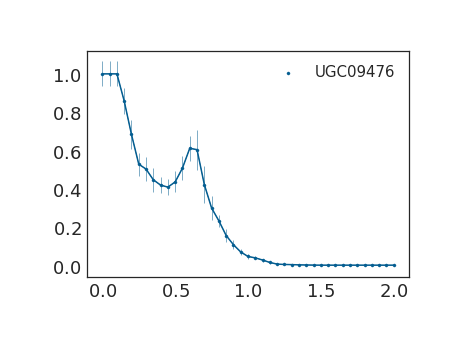}
\includegraphics[width=0.411\columnwidth,angle=0]{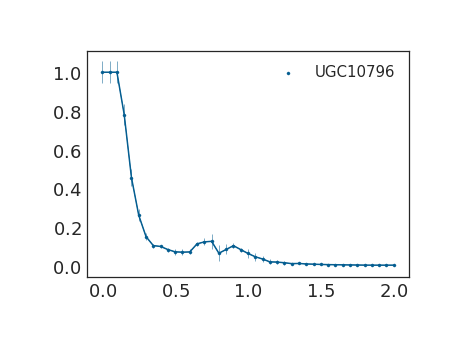}
\includegraphics[width=0.411\columnwidth,angle=0]{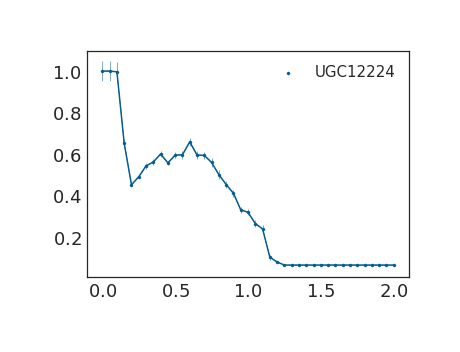}
\includegraphics[width=0.411\columnwidth,angle=0]{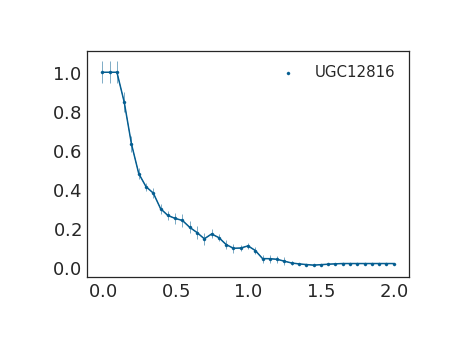}
\caption[]{CL Profiles, in units of $10^{-16} erg s^{-1} cm^{-2}$ for the H$\alpha$ emission.}
\label{CL_profiles}
\end{center}
\end{figure*}

\begin{figure*}
\begin{center}
\includegraphics[width=0.411\columnwidth,angle=0]{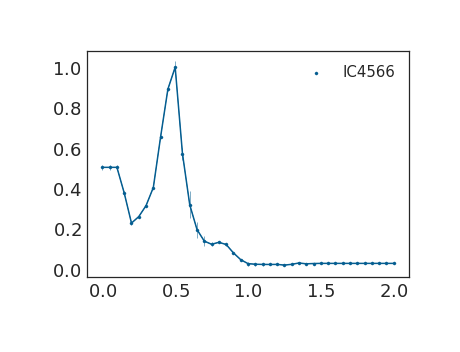}
\includegraphics[width=0.411\columnwidth,angle=0]{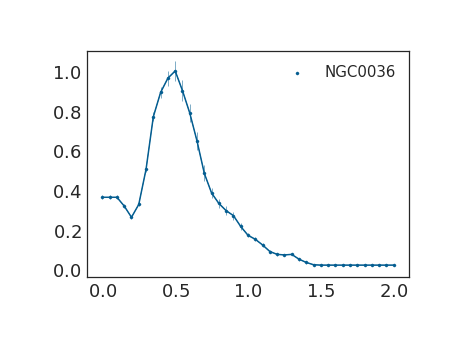}
\includegraphics[width=0.411\columnwidth,angle=0]{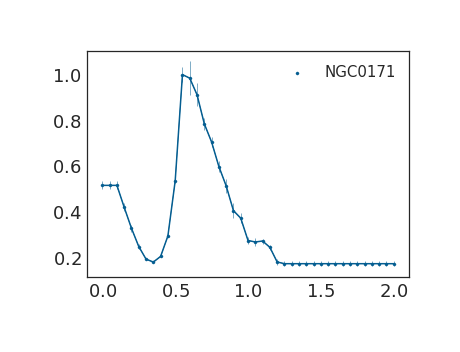}
\includegraphics[width=0.411\columnwidth,angle=0]{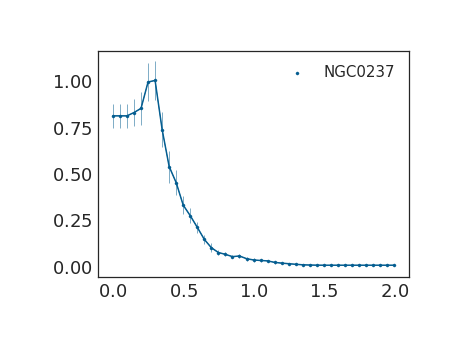}
\includegraphics[width=0.411\columnwidth,angle=0]{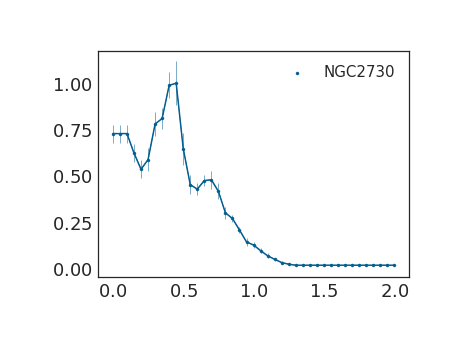}
\includegraphics[width=0.411\columnwidth,angle=0]{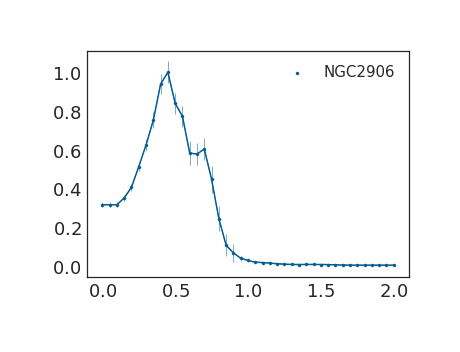}
\includegraphics[width=0.411\columnwidth,angle=0]{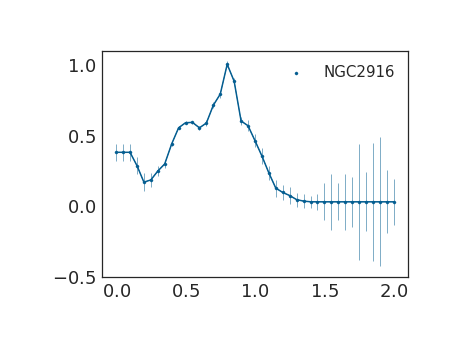}
\includegraphics[width=0.411\columnwidth,angle=0]{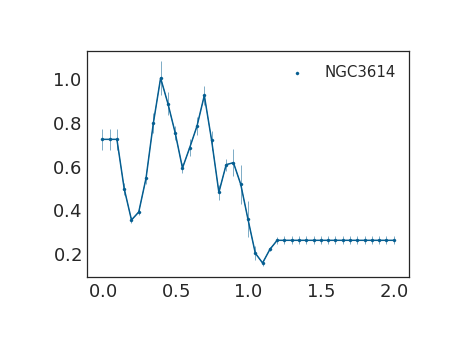}
\includegraphics[width=0.411\columnwidth,angle=0]{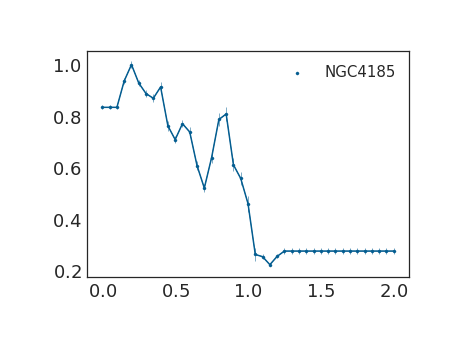}
\includegraphics[width=0.411\columnwidth,angle=0]{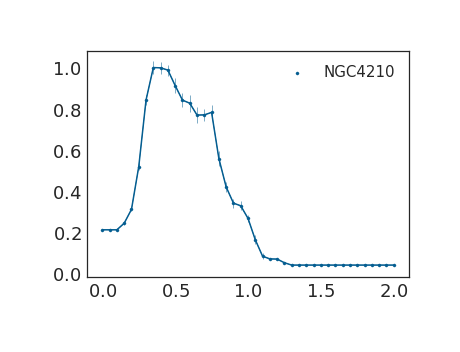}
\includegraphics[width=0.411\columnwidth,angle=0]{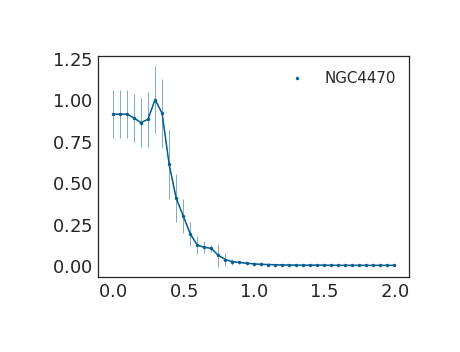}
\includegraphics[width=0.411\columnwidth,angle=0]{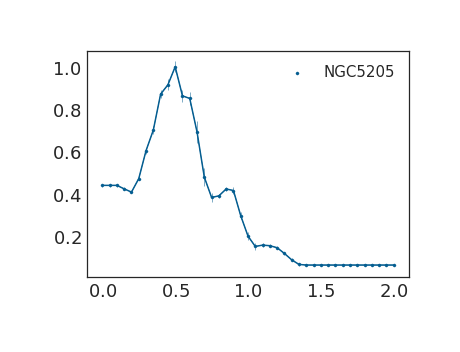}
\includegraphics[width=0.411\columnwidth,angle=0]{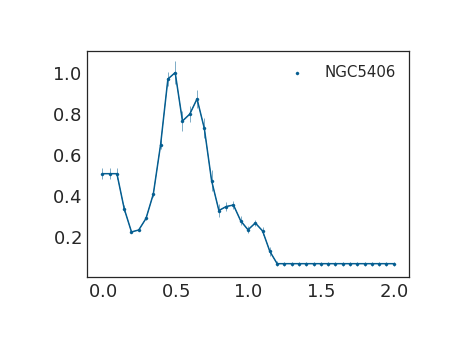}
\includegraphics[width=0.411\columnwidth,angle=0]{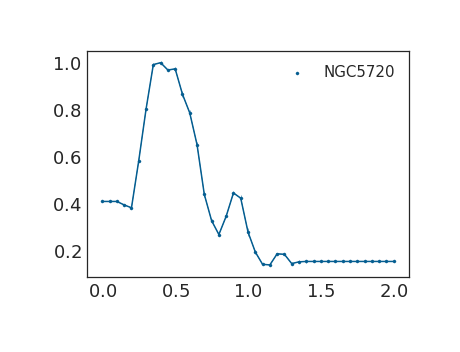}
\includegraphics[width=0.411\columnwidth,angle=0]{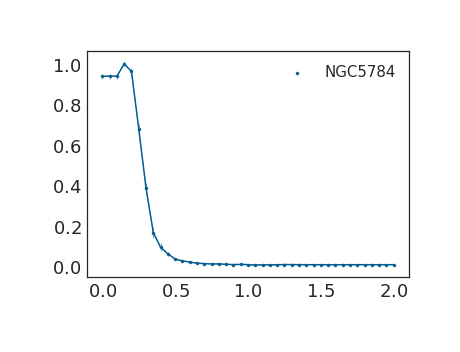}
\includegraphics[width=0.411\columnwidth,angle=0]{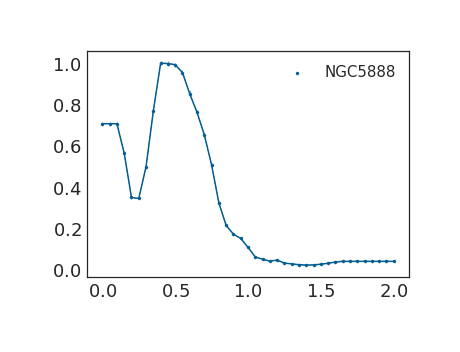}
\includegraphics[width=0.411\columnwidth,angle=0]{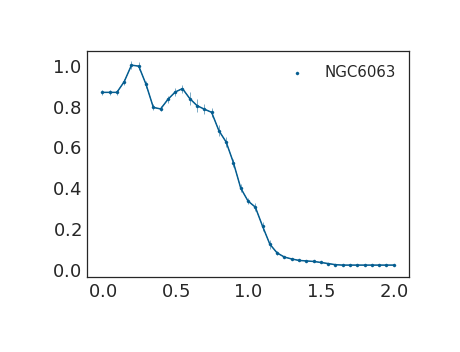}
\includegraphics[width=0.411\columnwidth,angle=0]{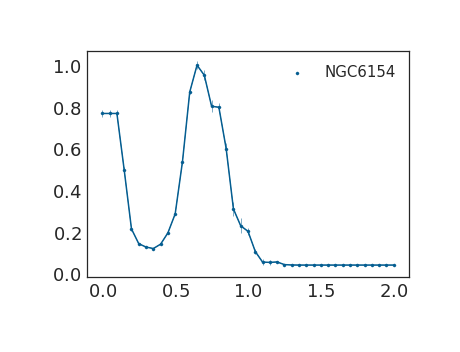}
\includegraphics[width=0.411\columnwidth,angle=0]{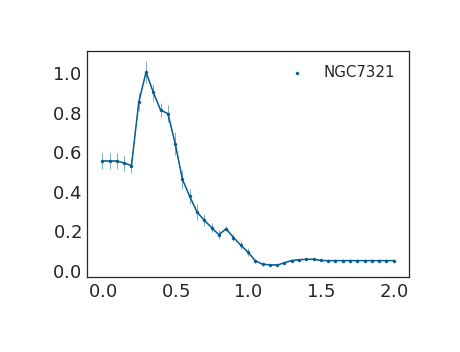}
\includegraphics[width=0.411\columnwidth,angle=0]{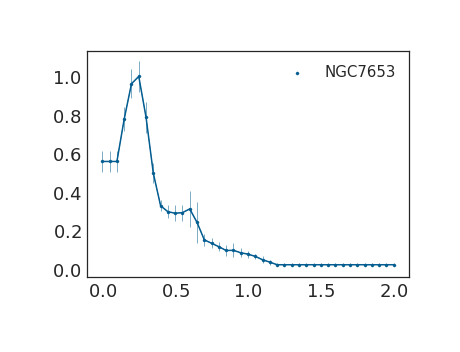}
\includegraphics[width=0.411\columnwidth,angle=0]{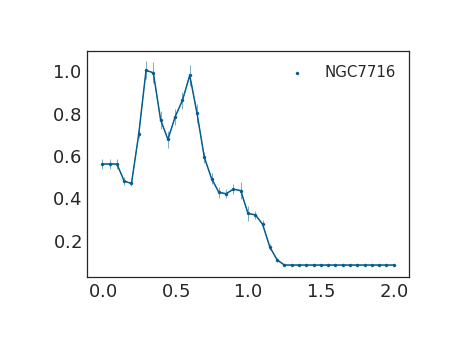}
\includegraphics[width=0.411\columnwidth,angle=0]{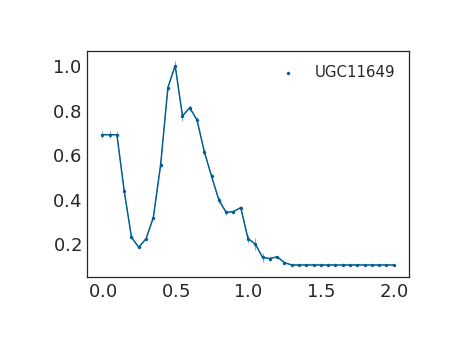}
\includegraphics[width=0.411\columnwidth,angle=0]{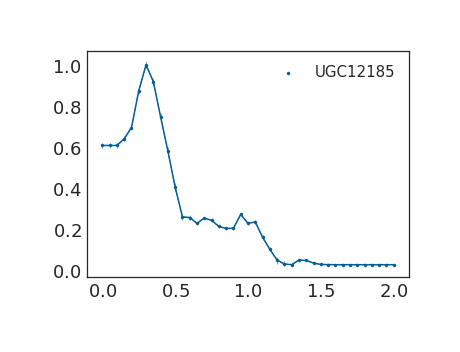}
\caption[]{EX Profiles, in units of $10^{-16} erg s^{-1} cm^{-2}$ for the H$\alpha$ emission.}
\label{E_profiles}
\end{center}
\end{figure*}

Figures \ref{CE_profiles}, \ref{CL_profiles} and \ref{E_profiles} shows the H$\alpha$ profiles for each galaxy in each profile type. The profiles are normalized by the H$\alpha$ maximum, in units of $10^{-16} erg s^{-1} cm^{-2}$.

\section{Galaxies Properties}
Table \ref{tab01} shows the list of galaxies in our sample and some global properties, like $Mr$ magnitude, u-r, mean stellar age, mean metallicity, mean stellar mass and Hubble Type. In addition, the table also shows the classification of the H$\alpha$ profile, the concentrations and the ellipticity for each object.
\onecolumn

\begin{center}
\begin{table}
\caption{Properties of the galaxies sample: absolute magnitude $M\underline{ }r$, color u-r, mean age ($log(age/yr)$), metallicity ($Z$), stellar mass ($log(M/M\underline{ }{\odot})$), Hubble type, morphological classes, H$\alpha$ profile, light concentration, effetive concentration,central concentration and ellipticity.}
\label{tab01}
\begin{tabular}{lcccccccccccc}
\hline
Galaxy & Mr  & u-r & log($t^*/Gy$) & log($Z^*/Z_{\odot}$)   & log($M_*/M_{\odot}$)  & H. Type     & M. Class     &  H$\alpha$ Profile &  $C_r$ & $C_E(H\alpha)$ & $C_C(H\alpha)$ & $\epsilon$\\
\hline
IC0776 & -18.69 & 1.94 & 8.34 & -0.27 & 9.59 & Sd & S\underline{ }late & CL & 1.95 & 0.55 & 0.20 & 0.28\\ 
IC1256 & -20.81 & 2.21 & 9.08 & -0.29 & 10.72 & Sb & S\underline{ }early & CL & 2.19 & 0.57 & 0.22 & 0.20\\ 
IC1683 & -20.75 & 2.54 & 9.31 & -0.24 & 10.76 & Sb & S\underline{ }early & CL & 2.59 & 0.30 & 0.77 & 0.28\\ 
IC4566 & -21.51 & 2.88 & 9.30 & -0.26 & 11.01 & Sb & S\underline{ }early & E & 2.24 & 0.63 & 0.08 & 0.26\\ 
NGC0001 & -21.11 & 2.24 & 8.89 & -0.35 & 10.82 & Sbc & S\underline{ }late & CE & 3.04 & 0.52 & 0.40 & 0.25\\ 
NGC0036 & -21.86 & 2.48 & 9.30 & -0.34 & 11.22 & Sb & S\underline{ }early & E & 2.49 & 0.72 & 0.05 & 0.25\\ 
NGC0171 & -21.24 & 2.59 & 9.18 & -0.21 & 10.93 & Sb & S\underline{ }early & E & 1.98 & 0.81 & 0.04 & 0.11\\ 
NGC0237 & -20.75 & 1.84 & 8.79 & -0.33 & 10.59 & Sc & S\underline{ }late & E & 2.42 & 0.55 & 0.33 & 0.27\\ 
NGC0477 & -20.79 & 2.39 & 8.89 & -0.33 & 10.89 & Sbc & S\underline{ }late & CL & 2.15 & 0.69 & 0.11 & 0.30\\ 
NGC0496 & -21.12 & 2.32 & 8.74 & -0.31 & 10.84 & Sc & S\underline{ }late & CL & 2.03 & 0.47 & 0.43 & 0.34\\ 
NGC0499 & -21.88 & 2.83 & 9.69 & -0.17 & 11.34 & E & E\underline{ }S0 & CE & 2.96 & 0.65 & 0.37 & 0.16\\ 
NGC0774 & -21.12 & 2.65 & 9.59 & -0.20 & 11.02 & S0 & E\underline{ }S0 & CE & 3.01 & 0.51 & 0.53 & 0.18\\ 
NGC0776 & -21.46 & 2.67 & 9.26 & -0.30 & 11.08 & Sb & S\underline{ }early & CL & 2.10 & 0.66 & 0.13 & 0.09\\ 
NGC0810 & -22.17 & 2.81 & 9.59 & -0.21 & 11.48 & E & E\underline{ }S0 & CE & 2.98 & 0.45 & 0.51 & 0.16\\ 
NGC1056 & -19.17 & 2.49 & 8.96 & -0.38 & 10.20 & Sa & S\underline{ }early & CE & 3.17 & 0.37 & 0.50 & 0.27\\ 
NGC1349 & -22.15 & 2.75 & 9.20 & -0.21 & 11.33 & E & E\underline{ }S0 & CL & 2.59 & 0.71 & 0.10 & 0.10\\ 
NGC1645 & -21.45 & 2.65 & 9.11 & -0.19 & 11.03 & S0 & E\underline{ }S0 & CE & 3.23 & 0.86 & 0.15 & 0.34\\ 
NGC2253 & -21.20 & 2.37 & 8.80 & -0.36 & 10.80 & Sbc & S\underline{ }late & CL & 2.14 & 0.66 & 0.13 & 0.22\\ 
NGC2347 & -21.60 & 2.15 & 8.64 & -0.33 & 11.04 & Sbc & S\underline{ }late & CL & 2.57 & 0.58 & 0.30 & 0.29\\ 
NGC2730 & -20.40 & 1.89 & 7.81 & -0.36 & 10.13 & Sc & S\underline{ }late & E & 1.88 & 0.62 & 0.11 & 0.20\\ 
NGC2880 & -20.24 & 2.65 & 9.70 & -0.25 & 10.55 & E & E\underline{ }S0 & CE & 3.26 & 0.81 & 0.17 & 0.21\\ 
NGC2906 & -20.56 & 2.96 & 9.42 & -0.31 & 10.58 & Sbc & S\underline{ }late & E & 2.08 & 0.64 & 0.17 & 0.38\\ 
NGC2916 & -21.17 & 2.18 & 8.94 & -0.31 & 10.96 & Sbc & S\underline{ }late & E & 2.16 & 0.87 & 0.02 & 0.31\\ 
NGC3106 & -22.16 & 2.96 & 9.49 & -0.23 & 11.40 & Sa & S\underline{ }early & CE & 3.06 & 0.84 & 0.12 & 0.10\\ 
NGC3381 & -19.32 & 1.84 & 8.81 & -0.33 & 9.88 & Sd & S\underline{ }late & CL & 2.03 & 0.59 & 0.20 & 0.07\\ 
NGC3614 & -19.96 & 2.52 & 9.22 & -0.42 & 10.31 & Sbc & S\underline{ }late & E & 2.18 & 0.77 & 0.04 & 0.10\\ 
NGC3811 & -20.76 & 2.30 & 8.93 & -0.35 & 10.64 & Sbc & S\underline{ }late & CL & 2.14 & 0.76 & 0.09 & 0.18\\ 
NGC4003 & -21.71 & 2.68 & 9.59 & -0.15 & 11.12 & S0 & E\underline{ }S0 & CE & 2.77 & 0.24 & 0.60 & 0.38\\ 
NGC4185 & -21.32 & 3.11 & 9.41 & -0.28 & 10.85 & Sbc & S\underline{ }late & E & 1.96 & 0.77 & 0.08 & 0.16\\ 
NGC4210 & -20.38 & 2.51 & 9.09 & -0.32 & 10.50 & Sb & S\underline{ }early & E & 1.87 & 0.73 & 0.03 & 0.18\\ 
NGC4470 & -20.44 & 1.50 & 9.01 & -0.38 & 10.23 & Sc & S\underline{ }late & E & 2.01 & 0.51 & 0.43 & 0.25\\ 
NGC4874 & -22.68 & 2.98 & 9.83 & -0.21 & 11.66 & E & E\underline{ }S0 & CE & 2.72 & 0.76 & 0.13 & 0.05\\ 
NGC5000 & -21.36 & 2.39 & 9.08 & -0.18 & 10.94 & Sbc & S\underline{ }late & CL & 2.19 & 0.70 & 0.17 & 0.13\\ 
NGC5205 & -19.34 & 2.38 & 8.99 & -0.33 & 9.98 & Sbc & S\underline{ }late & E & 2.42 & 0.74 & 0.06 & 0.25\\ 
NGC5216 & -20.28 & 2.66 & 9.56 & -0.22 & 10.58 & E & E\underline{ }S0 & CE & 2.95 & 0.43 & 0.52 & 0.12\\ 
NGC5378 & -20.60 & 2.85 & 9.48 & -0.27 & 10.61 & Sb & S\underline{ }early & CL & 2.42 & 0.86 & 0.09 & 0.15\\ 
NGC5406 & -22.03 & 2.81 & 9.33 & -0.29 & 11.26 & Sb & S\underline{ }early & E & 2.31 & 0.79 & 0.03 & 0.10\\ 
NGC5485 & -20.72 & 2.78 & 9.78 & -0.20 & 10.74 & E & E\underline{ }S0 & CE & 2.79 & 0.50 & 0.38 & 0.13\\ 
NGC5520 & -19.50 & 2.24 & 8.71 & -0.33 & 10.07 & Sbc & S\underline{ }late & CE & 3.03 & 0.52 & 0.45 & 0.37\\ 
NGC5614 & -22.04 & 2.79 & 9.61 & -0.37 & 11.22 & Sa & S\underline{ }early & CE & 3.05 & 0.71 & 0.19 & 0.08\\ 
NGC5720 & -21.90 & 2.40 & 9.12 & -0.32 & 11.19 & Sbc & S\underline{ }late & E & 2.36 & 0.73 & 0.06 & 0.24\\ 
NGC5784 & -21.96 & 2.84 & 9.56 & -0.29 & 11.32 & S0 & E\underline{ }S0 & E & 3.36 & 0.36 & 0.73 & 0.13\\ 
NGC5888 & -22.51 & 2.82 & 9.44 & -0.22 & 11.47 & Sb & S\underline{ }early & E & 2.07 & 0.69 & 0.08 & 0.36\\ 
NGC5966 & -21.43 & 2.77 & 9.70 & -0.25 & 11.12 & E & E\underline{ }S0 & CE & 2.83 & 0.59 & 0.43 & 0.22\\ 
NGC6004 & -21.14 & 3.46 & 9.29 & -0.34 & 10.86 & Sbc & S\underline{ }late & CL & 1.97 & 0.61 & 0.10 & 0.05\\ 
NGC6020 & -21.29 & 2.98 & 9.59 & -0.26 & 11.04 & E & E\underline{ }S0 & CE & 3.03 & 0.88 & 0.16 & 0.13\\ 
NGC6032 & -20.72 & 2.99 & 8.94 & -0.20 & 10.65 & Sbc & S\underline{ }late & CL & 1.96 & 0.64 & 0.27 & 0.44\\ 
NGC6063 & -20.02 & 2.35 & 9.04 & -0.38 & 10.36 & Sbc & S\underline{ }late & E & 1.89 & 0.74 & 0.14 & 0.36\\ 
NGC6125 & -22.01 & 2.74 & 9.78 & -0.24 & 11.35 & E & E\underline{ }S0 & CE & 3.19 & 0.80 & 0.14 & 0.06\\ 
NGC6146 & -22.94 & 2.73 & 9.59 & -0.09 & 11.71 & E & E\underline{ }S0 & CE & 3.36 & 0.50 & 0.47 & 0.14\\ 
NGC6150 & -22.28 & 2.86 & 9.65 & -0.15 & 11.42 & E & E\underline{ }S0 & CE & 2.91 & 1.01 & 0.33 & 0.29\\ 
\hline
\end{tabular}
\end{table}

\begin{table}
\contcaption{Properties of the galaxies sample: absolute magnitude $M\underline{ }r$, color u-r, mean age ($log(age/yr)$), metallicity ($Z$), stellar mass ($log(M/M\underline{ }{\odot})$), Hubble type, morphological classes, H$\alpha$ profile, light concentration, effetive concentration,central concentration and ellipticity.}
\begin{tabular}{lcccccccccccc}
\hline
Galaxy & Mr  & u-r & log($t^*/Gy$) & log($Z^*/Z_{\odot}$)   & log($M_*/M_{\odot}$)  & H. Type     & M. Class     &  H$\alpha$ Profile &  $C_r$ & $C_E(H\alpha)$ & $C_C(H\alpha)$ & $\epsilon$\\
\hline
NGC6154 & -21.60 & 2.73 & 9.48 & -0.28 & 11.14 & Sa & S\underline{ }early & E & 2.27 & 0.83 & 0.03 & 0.22\\ 
NGC6173 & -23.11 & 2.91 & 9.81 & -0.22 & 11.81 & E & E\underline{ }S0 & CE & 3.15 & 0.87 & 0.15 & 0.21\\ 
NGC6338 & -22.76 & 3.18 & 9.72 & -0.22 & 11.67 & E & E\underline{ }S0 & CE & 2.81 & 0.58 & 0.37 & 0.17\\ 
NGC6411 & -21.61 & 3.10 & 9.66 & -0.21 & 11.15 & E & E\underline{ }S0 & CE & 3.00 & 0.76 & 0.22 & 0.22\\ 
NGC6515 & -22.11 & 2.73 & 9.44 & -0.26 & 11.37 & E & E\underline{ }S0 & CE & 3.13 & 0.78 & 0.26 & 0.16\\ 
NGC6941 & -21.87 & 2.76 & 9.27 & -0.28 & 11.21 & Sb & S\underline{ }early & CL & 2.12 & 0.80 & 0.07 & 0.26\\ 
NGC7194 & -22.35 & 2.79 & 9.67 & -0.13 & 11.56 & E & E\underline{ }S0 & CE & 3.27 & 0.88 & 0.24 & 0.29\\ 
NGC7321 & -22.07 & 2.48 & 9.04 & -0.32 & 11.30 & Sbc & S\underline{ }late & E & 2.39 & 0.63 & 0.18 & 0.28\\ 
NGC7489 & -21.47 & 1.86 & 8.03 & -0.27 & 11.18 & Sbc & S\underline{ }late & CL & 2.28 & 0.66 & 0.15 & 0.14\\ 
NGC7562 & -21.88 & 2.86 & 9.76 & -0.15 & 11.31 & E & E\underline{ }S0 & CE & 3.07 & 0.75 & 0.21 & 0.21\\ 
NGC7563 & -21.16 & 2.99 & 9.74 & -0.03 & 11.13 & Sa & S\underline{ }early & CE & 3.52 & 0.93 & 0.20 & 0.27\\ 
NGC7591 & -21.31 & 2.38 & 8.99 & -0.33 & 10.95 & Sbc & S\underline{ }late & CE & 2.97 & 0.53 & 0.28 & 0.33\\ 
NGC7625 & -19.12 & 2.37 & 9.14 & -0.41 & 10.22 & Sa & S\underline{ }early & CL & 2.61 & 0.37 & 0.48 & 0.12\\ 
NGC7653 & -21.11 & 2.12 & 8.84 & -0.40 & 10.82 & Sb & S\underline{ }early & E & 2.73 & 0.64 & 0.18 & 0.12\\ 
NGC7671 & -21.28 & 2.73 & 9.40 & -0.23 & 11.03 & S0 & E\underline{ }S0 & CE & 3.22 & 0.63 & 0.38 & 0.25\\ 
NGC7683 & -21.14 & 2.89 & 9.66 & -0.21 & 11.01 & S0 & E\underline{ }S0 & CE & 3.16 & 0.46 & 0.50 & 0.29\\ 
NGC7716 & -20.29 & 2.49 & 9.13 & -0.37 & 10.65 & Sb & S\underline{ }early & E & 2.73 & 0.81 & 0.05 & 0.12\\ 
NGC7800 & -18.37 & 1.41 & 8.49 & -0.33 & 9.67 & Sd & S\underline{ }late & CL & 2.44 & 0.52 & 0.31 & 0.35\\ 
NGC7819 & -20.57 & 2.30 & 8.68 & -0.29 & 10.61 & Sc & S\underline{ }late & CL & 2.11 & 0.57 & 0.39 & 0.29\\ 
UGC03995 & -21.72 & 3.37 & 9.29 & -0.30 & 11.15 & Sb & S\underline{ }early & CL & 2.43 & 0.73 & 0.12 & 0.33\\ 
UGC05771 & -21.91 & 2.72 & 9.61 & -0.24 & 11.27 & E & E\underline{ }S0 & CE & 3.35 & 0.46 & 0.57 & 0.22\\ 
UGC07012 & -19.34 & 1.42 & 8.41 & -0.26 & 9.90 & Sc & S\underline{ }late & CE & 2.72 & 0.48 & 0.45 & 0.33\\ 
UGC08234 & -22.41 & 2.35 & 9.10 & -0.17 & 11.39 & S0 & E\underline{ }S0 & CE & 3.32 & 0.83 & 0.28 & 0.30\\ 
UGC08733 & -18.73 & 1.63 & 8.75 & -0.27 & 9.61 & Sd & S\underline{ }late & CL & 2.14 & 0.75 & 0.12 & 0.22\\ 
UGC08781 & -22.09 & 2.69 & 9.29 & -0.19 & 11.27 & Sb & S\underline{ }early & CL & 2.47 & 0.89 & 0.06 & 0.33\\ 
UGC09476 & -20.35 & 2.09 & 9.02 & -0.36 & 10.43 & Sbc & S\underline{ }late & CL & 1.84 & 0.65 & 0.19 & 0.26\\ 
UGC10693 & -22.66 & 2.94 & 9.74 & -0.28 & 11.61 & E & E\underline{ }S0 & CE & 3.28 & 0.77 & 0.19 & 0.13\\ 
UGC10695 & -22.09 & 2.73 & 9.57 & -0.27 & 11.38 & E & E\underline{ }S0 & CE & 2.80 & 0.88 & 0.31 & 0.37\\ 
UGC10796 & -18.90 & 1.75 & 8.54 & -0.23 & 9.77 & Sc & S\underline{ }late & CL & 2.41 & 0.63 & 0.42 & 0.34\\ 
UGC10905 & -22.32 & 2.78 & 9.53 & -0.26 & 11.45 & S0 & E\underline{ }S0 & CE & 3.47 & 0.48 & 0.52 & 0.35\\ 
UGC11649 & -20.60 & 2.88 & 9.22 & -0.22 & 10.75 & Sa & S\underline{ }early & E & 2.12 & 0.79 & 0.04 & 0.15\\ 
UGC12127 & -22.49 & 3.31 & 9.71 & -0.31 & 11.65 & E & E\underline{ }S0 & CE & 3.05 & 0.80 & 0.15 & 0.09\\ 
UGC12185 & -21.26 & 2.60 & 9.14 & -0.25 & 10.99 & Sb & S\underline{ }early & E & 2.83 & 0.68 & 0.25 & 0.38\\ 
UGC12224 & -19.92 & 2.51 & 8.83 & -0.27 & 10.41 & Sc & S\underline{ }late & CL & 1.84 & 0.74 & 0.09 & 0.09\\ 
UGC12816 & -20.28 & 1.85 & 8.38 & -0.23 & 10.34 & Sc & S\underline{ }late & CL & 2.37 & 0.63 & 0.31 & 0.32\\ 
\hline
\end{tabular}
\end{table}

\end{center}

\bsp	
\label{lastpage}

\end{document}